\documentclass[journal,twoside,web]{ieeecolor}
\usepackage{generic}
\usepackage{cite}
\usepackage{amsmath,amssymb,amsfonts}
\usepackage{algorithmic}
\usepackage{graphicx}
\usepackage{algorithm,algorithmic}
\usepackage{hyperref}
\hypersetup{hidelinks=true}
\usepackage{textcomp}
\usepackage{booktabs}
\usepackage{array}
\usepackage{colortbl}%
\usepackage{multirow}
\usepackage{booktabs}
\usepackage{bbm, dsfont}
\usepackage[table]{xcolor}
  \newcommand{\myrowcolour}{\rowcolor[gray]{0.925}}
\newcolumntype{C}[1]{>{\centering\arraybackslash}m{#1}}
\def\BibTeX{{\rm B\kern-.05em{\sc i\kern-.025em b}\kern-.08em
    T\kern-.1667em\lower.7ex\hbox{E}\kern-.125emX}}
\begin{document}
\title{MetaPlate: Counterfactual-Guided RAG-LLM Personalized Food Recommendation for Hyperglycemia Prevention}
\author{Asiful Arefeen\textsuperscript{1,2}*, Carol Johnston\textsuperscript{1}, Hassan Ghasemzadeh\textsuperscript{1}
\thanks{\textsuperscript{1}College of Health Solutions, Arizona State University, Phoenix, AZ 85004, USA}
\thanks{\textsuperscript{2}School of Computing and Augmented Intelligence, Arizona State University, Tempe, AZ 85281, USA}
\thanks{*\textbf{Corresponding author}: aarefeen@asu.edu}
\thanks{**\textbf{Code available at:} \href{https://github.com/Arefeen06088/MetaPlate}{\textcolor{blue}{github.com/Arefeen06088/MetaPlate}}}
}

\maketitle

\begin{abstract}
Postprandial hyperglycemia is a key risk factor for metabolic disorders; however, existing dietary guidance is often static, impractical, and insufficiently personalized, frequently providing low-level recommendations that are difficult to follow or high-level recommendations that are not impactful. While recent advances leverage continuous glucose monitoring (CGM) and machine learning to predict glycemic responses, these approaches are predominantly predictive and lack actionable dietary guidance. Moreover, existing recommendation systems are often insufficiently aligned with user goals and may require extensive user input. To address these gaps in research, we present \textbf{\textit{MetaPlate}}, a counterfactual explanation (CF)-guided, context-aware decision-support framework that generates personalized meal recommendations to mitigate postprandial glucose excursions. MetaPlate integrates multimodal data, including CGM readings, wearable-derived physiological signals, and user-provided meal inputs to model pre-meal context. A machine learning model predicts postprandial glucose response for a given meal and context (RMSE of $16.46$ mg/dL), while a CF optimization module adjusts meal composition—modifying macro-nutrient amounts—to maintain glucose levels within a target range (i.e., $\leq$$140$ mg/dL). An LLM-based retrieval-augmented generation (RAG) layer further enhances interpretability and understanding by producing human-readable meal recommendations and explanations utilizing constrained search from the USDA food database. We first evaluate MetaPlate using data collected in an observational study involving $25$ healthy adults. MetaPlate achieves higher validity ($0.660$ vs. $0.540$ for other techniques) while requiring smaller meal modifications (normalized L1 distance: $0.230$). We then assess MetaPlate through a structured, expert-in-the-loop assessment involving registered dietitians (RDs), comparing system performance before and after prompt refinement. Results demonstrate substantial improvements in portion suitability (from $4.88$ to $7.67$), alignment with dietary guidelines (from $5.38$ to $7.97$), and recommendation likelihood (from $3.45$ to $7.10$) on a 10-point Likert scale, with expert feedback indicating a shift from clinically implausible outputs toward actionable and contextually appropriate recommendations. Our findings emphasizes the importance of adding domain knowledge and structured constraints in LLM-driven systems and underscore the potential of MetaPlate as a real-time, personalized dietary decision-support tool for hyperglycemia prevention.
\end{abstract}


\section{Introduction}
\label{sec:introduction}
\IEEEPARstart{P}{ostprandial} hyperglycemia is defined as blood glucose concentrations exceeding $140$ mg/dL in healthy individuals \cite{Jarvis2023ContinuousGM}. Persistent hyperglycemia contributes to the onset of diabetes mellitus \cite{Giri2018ChronicHM} and is associated with elevated cardiovascular risk \cite{Levitan2004IsNH}. Therefore, maintaining in-range postprandial glucose levels is important not only for individuals with diabetes but also for otherwise healthy adults. Notably, recent studies using continuous glucose monitoring (CGM) have demonstrated substantial inter-individual variability in glycemic responses to identical meals, exposing the limitations of generalized, one-size-fits-all dietary guidelines \cite{Phillips2023UncoveringPG}. These risk factors, together with the challenges associated with glucose variability, emphasize the need for personalized, context-aware dietary strategies to mitigate glycemic excursions in healthy adults prior to the onset of overt metabolic disease.

Recent work has increasingly leveraged CGM and wearable or physiological sensing to model and forecast glycemic response. Seminal work by Zeevi et al. demonstrated that glycemic responses to real-life meals can be predicted using machine learning models that combine meal characteristics with individual-level biological and behavioral features \cite{Zeevi2015PersonalizedNB}. Subsequent studies have extended this direction by integrating recent CGM traces, activity signals, and multimodal wearable data to predict postprandial glucose response or hyperglycemia risk \cite{vanDoorn2021MachineLG, Farahmand2025AttenGlucoMT, Machiraju2025TimeAwareCF, Zhu2024PopulationSpecificGP, Shroff2023GlucoseAssistPB, Khamesian2025GlycemicAwareAA, Soumma2026GlyRAGCR}. However, these systems largely stop at forecasting: they estimate \textit{what is likely to happen after a meal}, but do not directly address the more actionable question of \textit{what a user should eat instead} under a given pre-meal context.

In parallel, food and nutrition recommendation systems have advanced rapidly from preference-based recommenders \cite{Kalpakoglou2025AnAN} toward goal-driven \cite{Arefeen2022ComputationalFF}, health-aware and RAG-enabled LLM frameworks \cite{Gavai2025AIdrivenPN} for meal planning tools. Recent approaches like NutriGen \cite{Khamesian2025NutriGenPM} leverages prompt engineering to generate personalized meal plans aligned with user preferences and constraints, while ChatDiet \cite{YANG2024100465} combines LLMs with population and person-level models to recommend meals that minimize glycemic excursions while improving explainability and interactivity. MOPI-HFRS \cite{Zhang2024MOPIHFRSAM} optimizes user preferences, personalized healthiness, and nutritional diversity to generate meal recommendations, followed by LLM-aided interpretation. Broader work in food recommendation has also emphasized the importance of personalization, contextual awareness, and health objectives. Despite these advances, most existing food recommendation systems remain centered on dietary preferences, calorie targets, or general wellness constraints rather than on immediate physiological state and explicit glycemic goals. They also often depend on extensive user-specified inputs and therefore less suitable for just-in-time decision support at mealtime.

CFs offer a promising paradigm for bridging the gap between prediction and intervention. Rather than merely explaining why a harmful outcome is predicted, CF methods identify minimal, feasible changes to inputs that would alter the outcome. In the dietary setting, this translates to the question: \textit{how should the meal be changed so that the predicted postprandial glucose response remains within a healthy range}? This framing is particularly well-suited for behavior change, because it translates model output into concrete, goal-directed suggestions. Prior work has highlighted the potential of CF not only as interpretability tools but also as a basis for actionable health interventions \cite{Crupi2021CounterfactualEA}.

Despite this promise, current CF-based methods for glycemic management still exhibit important limitations. Existing approaches like ExAct \cite{Arefeen2023DesigningUB}, GlyMan, and GlyTwin \cite{Arefeen2025GlyTwinDT} primarily generate interventions at the level of behavioral variables or macronutrient adjustments, like changing carbohydrate intake \cite{Arefeen2023DesigningUB, Arefeen2025GlyTwinDT}, insulin dosage \cite{Arefeen2025GlyTwinDT, Arefeen2024GlyManGM}, or other numerical treatment parameters \cite{Arefeen2023DesigningUB}. While useful, these low-level recommendations do not fully realize the practical potential of explainable AI in everyday dietary decision-making \cite{Barbiero2021EntropybasedLE}, as users ultimately make choices in terms of foods and portion sizes rather than abstract nutrient quantities. As a result, the final step from CF reasoning to human-understandable meal guidance often remains underdeveloped.

To address these gaps, we present MetaPlate, a CF-guided, context-aware decision-support framework for personalized meal recommendation in healthy adults. MetaPlate integrates multimodal pre-meal information from CGM, wearable-derived physiological signals, and user meal input to model an individual’s current context. A machine learning model first predicts the postprandial glucose response of a planned meal; a CF optimization module then adjusts meal macronutrient composition to satisfy a glycemic target; finally, an LLM-based retrieval-augmented layer grounds the recommendation in the USDA food database and converts these optimized nutritional targets into food-level, human-readable meal suggestions. In addition, we integrate an expert-in-the-loop refinement process to improve the clinical plausibility and usability of generated meals, ensuring that recommendations are not only glycemically effective but also realistic and actionable in practice. Thus, MetaPlate moves beyond passive forecasting and low-level CF outputs to deliver interpretable, user-centered, and goal-aligned dietary guidance.

We make the following contributions:
\begin{itemize}
\item We introduce a \textbf{context-aware multimodal framework} that combines CGM, wearable sensing, and user meal input to support pre-meal glycemic decision-making in healthy adults.

\item We formulate meal recommendation as a \textbf{CF intervention problem} and program an optimization system to generate meal modifications targeted to maintain postprandial glucose at or below a desired threshold.

\item We bridge the gap between numeric crude level CF outputs and practical dietary guidance through an \textbf{LLM-based RAG layer} that maps optimized macronutrient targets to food-level recommendations using the USDA database.

\item We introduce an \textbf{expert-in-the-loop evaluation and refinement pipeline} realizing that domain knowledge significantly improves meal realism, portion suitability, and clinical usability of LLM-generated recommendations.

\item We evaluate the practical validity of MetaPlate using \textbf{a new dataset collected in free-living condition} and \textbf{structured expert assessment}, focusing on clinical appropriateness, nutritional soundness, and practical utility after iterative refinement.
\end{itemize}

\section{Problem Statement}
\label{sec:problem}

Let $\mathcal{D} = \{(\mathbf{x}_i, \mathbf{m}_i, g_i)\}_{i=1}^{n}$ denote a dataset of $n$ eating events collected from individuals, where each instance consists of a pre-meal context $\mathbf{x}_i \in \mathbb{R}^d$, a meal representation $\mathbf{m}_i \in \mathbb{R}^k$, and the corresponding observed postprandial glucose outcome $g_i \in \mathbb{R}$ (e.g., peak glucose within a specified post-meal window).

The pre-meal context vector $\mathbf{x}_i = [x_i^1, x_i^2, \dots, x_i^d]$ includes multimodal physiological and behavioral features derived from CGM and wearable sensors. These features may include recent glucose trends, activity levels, heart rate, and other contextual signals. The meal vector $\mathbf{m}_i = [m_i^1, m_i^2, \dots, m_i^k]$ encodes the nutritional composition of the meal, such as macronutrient quantities (e.g., carbohydrates, protein, and fat).

We assume a predictive function
\begin{equation}
f: \mathbb{R}^{d} \times \mathbb{R}^{k} \rightarrow \mathbb{R},
\end{equation}
where $\mathbb{R}^d \times \mathbb{R}^k$ denotes the joint input space of context and meal features, which are concatenated in practice. The function $f$ maps a pre-meal context $\mathbf{x}$ and a meal $\mathbf{m}$ to a predicted postprandial glucose response, i.e.,
\begin{equation}
f(\mathbf{x}, \mathbf{m}) = \hat{g}
\end{equation}
Given a target glycemic threshold $\tau$ (e.g., $\tau = 140$ mg/dL), the desired outcome is:
\begin{equation}
f(\mathbf{x}, \mathbf{m}) \leq \tau.
\end{equation}

However, for a planned meal $\mathbf{m}_0$, it is often observed that:
\begin{equation}
f(\mathbf{x}, \mathbf{m}_0) > \tau,
\end{equation}
indicating a potential hyperglycemic response.

The objective is therefore to identify a modified meal $\mathbf{m}^*$ that minimizes deviation from the original meal while satisfying the glycemic constraint:
\begin{equation}
\mathbf{m}^* = \arg\min_{\mathbf{m}} \; \mathcal{D}(\mathbf{m}, \mathbf{m}_0)
\end{equation}
subject to:
\begin{equation}
f(\mathbf{x}, \mathbf{m}) \leq \tau, \quad \mathbf{m} \in \mathcal{M},
\end{equation}
where $\mathcal{D}(\cdot)$ is a distance function measuring the magnitude of modification (e.g., changes in macronutrient composition), and $\mathcal{M}$ denotes the space of feasible meals under nutritional, dietary, and practical constraints.

Additionally, we distinguish between actionable and non-actionable components within the input space. While the pre-meal context $\mathbf{x}$ contains largely non-actionable physiological signals, the meal vector $\mathbf{m}$ represents the primary actionable component. Therefore, the intervention is constrained to modifying $\mathbf{m}$ while keeping $\mathbf{x}$ fixed.

In practice, the optimized meal $\mathbf{m}^*$ must be realizable as a combination of discrete food items and portion sizes. Thus, beyond satisfying the optimization constraints, the modified meal $\mathbf{m}^*$ should be:
(i) nutritionally valid,
(ii) practically realizable as a combination of real food items and portion sizes, and
(iii) interpretable to the end user.

In summary, the problem can be formulated as a constrained counterfactual optimization task: \textit{given a pre-meal context $\mathbf{x}$ and an initial meal $\mathbf{m}_0$, can we generate a minimally modified, feasible meal $\mathbf{m}^*$ that achieves the desired glycemic outcome while remaining actionable and interpretable?}

\section{Proposed Method}
\label{sec:method}

\subsection{Postprandial Glucose Forecasting Model}

Given a pre-meal context $\mathbf{x} \in \mathbb{R}^d$ and a meal representation $\mathbf{m} \in \mathbb{R}^k$, the goal of the forecasting model is to predict the corresponding postprandial glucose response.

We learn a function
\begin{equation}
f_\theta: \mathbb{R}^{d} \times \mathbb{R}^{k} \rightarrow \mathbb{R},
\end{equation}
parameterized by $\theta$, that estimates the peak postprandial glucose level:
\begin{equation}
\hat{g} = f_\theta(\mathbf{x}, \mathbf{m}).
\end{equation}

In practice, the inputs $\mathbf{x}$ and $\mathbf{m}$ are concatenated into a joint feature representation and passed through a model that captures temporal and contextual dependencies. The model is trained on the dataset $\mathcal{D} = \{(\mathbf{x}_i, \mathbf{m}_i, g_i)\}_{i=1}^n$ by minimizing a regression loss:
\begin{equation}
\mathcal{L}_{pred} = \frac{1}{n} \sum_{i=1}^{n} \left( f_\theta(\mathbf{x}_i, \mathbf{m}_i) - g_i \right)^2.
\end{equation}

This predictive model serves as the foundation for downstream counterfactual optimization.

\subsection{Counterfactual Optimization for Meal Adjustment}

Given a pre-meal context $\mathbf{x}$ and an initial planned meal $\mathbf{m}_0$, our objective is to generate a modified meal $\mathbf{m}^*$ such that the predicted glucose response satisfies the target constraint $f_\theta(\mathbf{x}, \mathbf{m}^*) \leq \tau$, while minimizing deviation from the original meal and ensuring it is practically realizable. This is formulated as a CF optimization problem, where the goal is to identify the minimal intervention on the actionable input $\mathbf{m}$ such that the predicted postprandial glucose response remains within a clinically acceptable range.

Inspired by prior work on counterfactual explanations and intervention modeling, we require the generated meal $\mathbf{m}^*$ to satisfy the following properties:
\begin{itemize}
    \item \textit{\textbf{Interventional}}: The modified meal must achieve the desired glycemic outcome, i.e., $f(\mathbf{x}, \mathbf{m}^*) \leq \tau$.
    \item \textit{\textbf{Minimal}}: The modification from the original meal $\mathbf{m}_0$ should be as small as possible.
    \item \textit{\textbf{Actionable}}: Only meal-related variables are modified, while the pre-meal context $\mathbf{x}$ remains fixed.
    \item \textit{\textbf{Plausible}}: The modified meal must satisfy nutritional and practical constraints, ensuring feasibility in real-world settings.
\end{itemize}

To implement the CF optimization framework, one straightforward option can be using an existing CF framework like DiCE \cite{mothilal2020dice}, NICE \cite{Brughmans2021NICEAA} or CFNOW \cite{DEOLIVEIRA2023}. However, to exert more control over the optimization (e.g. direction and limit of changes) and to generate realistic looking CFs (e.g. not suggesting to increase carbohydrate intake to reduce post-meal BGL), we choose to design our own optimization method. We formulate this as a constrained optimization problem using a differentiable objective function:
\begin{equation}
\label{eq:cf_loss}
\mathcal{L}(\mathbf{m}) =
\underbrace{\left[\max\left(0, f(\mathbf{x}, \mathbf{m}) - \tau \right)\right]^2}_{\text{target loss}}
+ 
\underbrace{\lambda \sum_{j \in \mathcal{A}}
\left(\frac{m_j - m_{0,j}}{s_j}\right)^2}_{\text{distance regularizer}},
\end{equation}

where the first term enforces the glycemic target by penalizing predictions that exceed the threshold $\tau$, and the second term is a distance regularizer that encourages minimal deviation from the original meal $\mathbf{m}_0$. Here, $\mathcal{A}$ denotes the set of actionable nutritional components (e.g., carbohydrates, protein, and fat), $s_j$ is a scaling factor for normalization across different nutrient ranges, and $\lambda$ controls the trade-off between achieving the glycemic target and preserving similarity to the original meal.

The optimization is subject to domain-specific constraints that ensure nutritional validity and practical feasibility:
\begin{equation}
\label{eq:constraints}
\begin{aligned}
m_C &\leq m_{0,C}, \\
m_P &\leq P_{\max}, \\
m_F &\leq F_{\max}, \\
\mathbf{m} &\in \mathcal{M},
\end{aligned}
\end{equation}
where $m_C$, $m_P$, and $m_F$ denote carbohydrate, protein, and fat components of the meal, respectively. The constraint $m_C \leq m_{0,C}$ reflects the dominant role of carbohydrate intake in postprandial glucose excursions, while $P_{\max}$ and $F_{\max}$ impose upper bounds to maintain nutritional balance. The feasible set $\mathcal{M}$ further ensures that the optimized meal adheres to dietary guidelines and practical consumption constraints.

The optimized solution is obtained as:
\begin{equation}
\mathbf{m}^* = \arg\min_{\mathbf{m}} \mathcal{L}(\mathbf{m}).
\end{equation}

This formulation differs from prior counterfactual approaches that operate on high-dimensional behavioral or clinical feature spaces and often produce low-level recommendations (e.g., adjusting macronutrient quantities or treatment parameters). In contrast, MetaPlate constrains the intervention space to meal composition and explicitly bridges the gap between numerical optimization and actionable dietary guidance. The optimized macronutrient targets obtained from this step are subsequently translated into food-level recommendations through an LLM-based retrieval and reasoning module.

\subsection{LLM-Based Retrieval-Augmented Food Recommendation}

\begin{figure*}[!h]
\centering
    \includegraphics[width=0.8\linewidth]{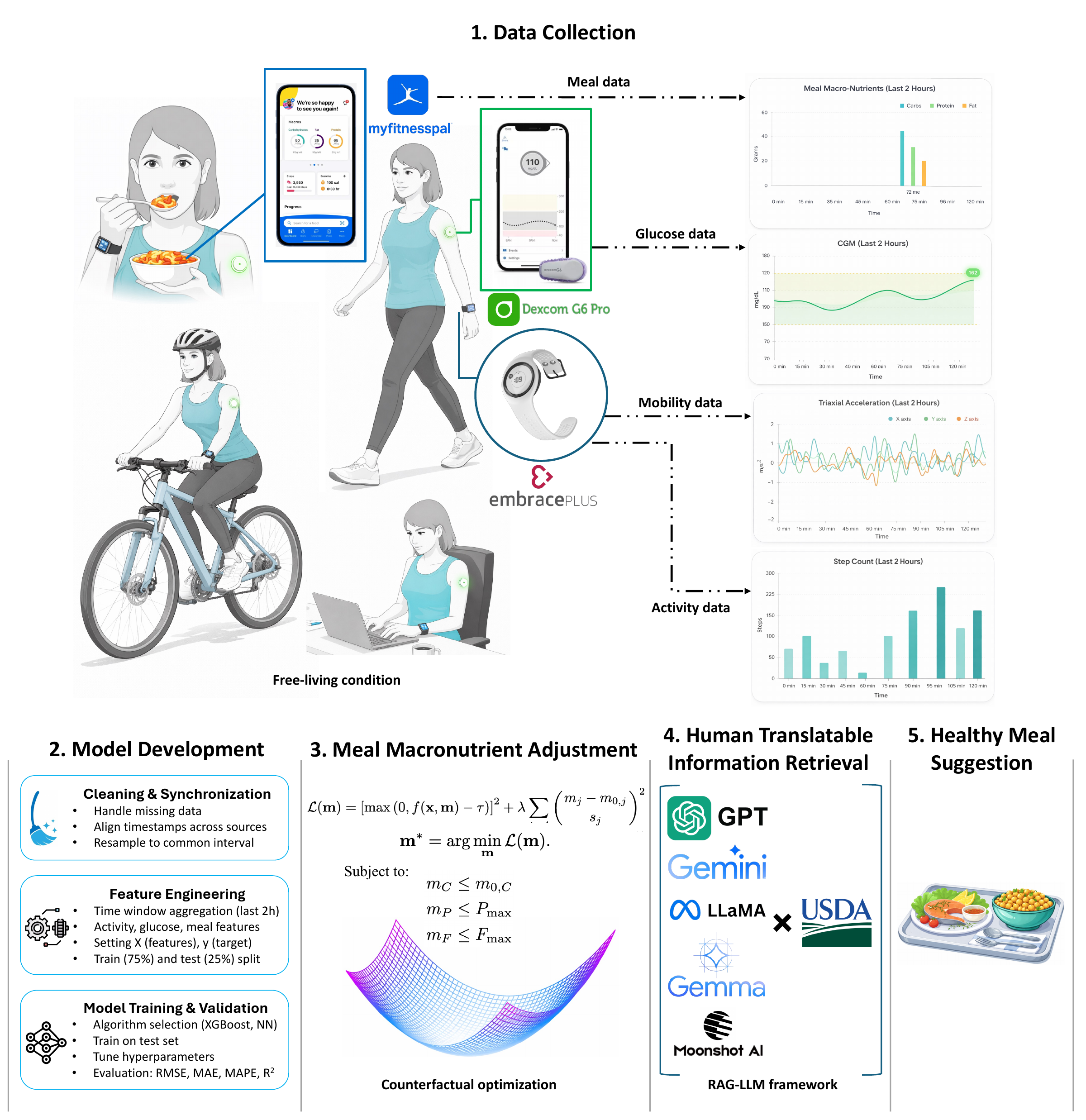}
\caption{MetaPlate framework consists of multiple phases: (1) data acquisition from healthy adults in free-living condition using CGM sensor, wristband and smartphone application, (2) data processing, feature engineering, model training and validation for forecasting post-meal blood glucose peak, (3) CF optimization for meal macro-nutrient adjustment to achieve in-range post-meal glucose level, (4) LLM-RAG module to retrieve human understandable healthy meals from USDA database satisfying constraints generated through CF optimization.}
\label{metaplate_diagram}
\end{figure*}

While the counterfactual optimization produces an adjusted macronutrient vector $\mathbf{m}^*$, these outputs are not directly interpretable by end users. To bridge this gap, we introduce a RAG layer that maps optimized nutritional targets to concrete food-level recommendations.

We construct a structured food database $\mathcal{F}$ derived from the USDA FoodData Central repository, where each food item is associated with its nutritional profile. Given the optimized macronutrient target $\mathbf{m}^*$, a retrieval module identifies candidate food combinations whose aggregate nutritional composition approximates $\mathbf{m}^*$.

Formally, the retrieval step identifies a set of food items $\{f_1, f_2, \dots, f_k\} \subset \mathcal{F}$ such that:
\begin{equation}
\sum_{j} \phi(f_j) \approx \mathbf{m}^*,
\end{equation}
where $\phi(f_j)$ denotes the nutrient vector of food item $f_j$.

An LLM is then used to generate human-readable meal recommendations conditioned on the retrieved candidates and user context. The LLM refines the output by adding dietary coherence, portion sizes, and natural language explanations, producing a final recommendation that is interpretable, actionable, and aligned with the optimized nutritional targets.

This RAG-based layer enables MetaPlate to translate low-level counterfactual outputs into practical dietary guidance suitable for real-world use.

\subsection{Prompt Optimization}

A baseline prompt was initially designed to generate meal recommendations that satisfy target macronutrient constraints using foods from the USDA FoodData Central database. The outputs produced by this prompt were reviewed by a cohort of domain experts with backgrounds in nutrition and dietetics.

Based on this evaluation, the prompt was iteratively refined to address identified limitations, including issues related to clinical plausibility and meal composition. Specifically, additional constraints were added to enforce balanced meal structure (e.g., inclusion of protein, carbohydrate, and non-starchy produce components), improve portion realism, and discourage implausible food combinations.

The finalized prompt, reflecting these refinements, is used in all subsequent experiments.

\subsection{Pipeline}

Figure~\ref{metaplate_diagram} presents an overview of the MetaPlate pipeline for personalized meal recommendation under glycemic constraints. The framework consists of five stages: data collection, data processing, model development, CF optimization, and LLM-based RAG recommendation.

In the data collection stage, multimodal data is obtained from healthy adults in free-living conditions using CGM, wearable sensors, and smartphone-based meal logs. These data include physiological signals (e.g., glucose, activity, heart rate) and meal information. The data processing stage performs cleaning, synchronization, missing value handling, and feature engineering to construct pre-meal context and meal representations.

In the model development stage, a predictive function is learned to estimate 2-hour postprandial glucose peak given the pre-meal context and meal. This model serves as the basis for subsequent optimization and is not restricted to a specific architecture.

The CF optimization stage modifies the planned meal to obtain a minimally changed alternative that satisfies a target glycemic constraint. The optimization is performed over macronutrient components under feasibility and nutritional constraints.

Finally, the LLM-based RAG module maps the optimized macronutrient targets to food-level recommendations by retrieving candidate items from the USDA database and generating interpretable meal suggestions with portioning and explanations.

\section{Data}

\subsection{Data Collection}
Data collection was conducted between May 2025 and April 2026 involving $13$ healthy individuals with no self-reported history of diabetes (IRB \#15102). All participants were familiar with meal logging. Following informed consent, participants were equipped with a Dexcom G6 Pro CGM sensor worn on the upper arm and an Embrace Plus wristband worn on the dominant wrist. The wristband was recharged every other night. Participants used the MyFitnessPal Calorie Tracker application on their smartphones to log dietary intake.

For each participant, approximately $10$ days of recording was collected under free-living conditions. The dataset includes continuous glucose measurements from the CGM, physiological signals (pulse rate, skin temperature, and electrodermal activity), and activity-related measures (step count, metabolic equivalents (METs), and activity counts) from the wristband, as well as meal timing and macro- and micronutrient information from food logs. In total, the dataset comprises approximately $3120$ hours of glucose data and $636$ logged meals.

Data from all 13 participants (age: $26.46 \pm 6.81$ years; 10 female; BMI: $26.23 \pm 5.65$) is used for subsequent analysis and model development. Table~\ref{cgm_data_2} summarizes participant demographics.

\begin{table}[h]
\footnotesize
\centering
\caption{Demographic information of the dataset collected from healthy adults.}
\label{cgm_data_2}
{\renewcommand{\arraystretch}{1.4}

\begin{tabular}{>{\raggedright\arraybackslash} C{0.08cm} C{1.3cm} C{0.8cm} C{1.3cm} C{1.3cm} C{1.3cm}}
\toprule
\textbf{\textit{n}} & \textbf{Age (mean$\pm$SD)} & \textbf{Gender (F/M)} & \textbf{Height (mean$\pm$SD) in.} & \textbf{Weight (mean$\pm$SD) lbs.} & \textbf{BMI (mean$\pm$SD)} \\
\myrowcolour%
\hline
13 &  26.46$\pm$6.81& 10/3 & 65.19$\pm$3.59 & 158.69$\pm$34.18 & 26.23$\pm$5.65 \\
\hline
\hline
\end{tabular}}
\end{table}

\begin{table*}[!h]
\caption{Examples of processed samples from the dataset. Pre BG slope, Pre BG and Post BG refer to Pre-meal blood glucose slope, Pre-meal blood glucose and 2h Post-meal max blood glucose, respectively. The modifiable features are marked with \colorbox{teal!40}{\phantom{a}}~teal color boxes.}
\label{sample}
\centering
\scriptsize
{\renewcommand{\arraystretch}{1.25}
\arrayrulecolor{black}

\begin{tabular}{|c|c|cc|cc|cc|cc|cc|c|c|c|c|c|c|}
\toprule
\multirow{2}{*}{\textbf{Age}} &
\multirow{2}{*}{\textbf{BMI}} &
\multicolumn{2}{c|}{\textbf{Step count}} &
\multicolumn{2}{c|}{\textbf{Activity count}} &
\multicolumn{2}{c|}{\textbf{EDA}} &
\multicolumn{2}{c|}{\textbf{Skin temp.}} &
\multicolumn{2}{c|}{\textbf{Pulse rate}} &
\multirow{2}{*}{\textbf{Pre BG slope}} &
\multirow{2}{*}{\textbf{Pre BG}} &
\multirow{2}{*}{\textbf{Carb}} &
\multirow{2}{*}{\textbf{Prot}} &
\multirow{2}{*}{\textbf{Fat}} &
\multirow{2}{*}{\textbf{Post BG}} \\
\cmidrule(lr){3-4}\cmidrule(lr){5-6}\cmidrule(lr){7-8}\cmidrule(lr){9-10}\cmidrule(lr){11-12}
& &
\textbf{mean} & \textbf{std} &
\textbf{mean} & \textbf{std} &
\textbf{mean} & \textbf{std} &
\textbf{mean} & \textbf{std} &
\textbf{mean} & \textbf{std} &
& & & & & \\
\hline
40 & 28.6 & 17.6 & 21.1 & 71.1 & 42.2 & 3.8 & 4.6 & 31.9 & 0.7 & 87.7 & 9.7 & -0.23 & 93 & \cellcolor{teal!25}52 & \cellcolor{teal!25}7 & \cellcolor{teal!25}4 & \textcolor{cyan}{\textbf{106}} \\
\hline
25 & 23.3 & 7.1 & 15.3 & 42.5 & 34.7 & 2.4 & 0.2 & 32.3 & 0.2 & 76.6 & 9.5 & 0.03 & 88 & \cellcolor{teal!25}74 & \cellcolor{teal!25}11.1 & \cellcolor{teal!25}7.5 & \textcolor{red}{\textbf{169}} \\
\hline
\bottomrule
\end{tabular}}
\end{table*}

\subsection{Data Processing}
For robust modeling of postprandial glycemic response under free-living conditions, a multi-stage preprocessing pipeline is developed to synchronize, clean and transform multimodal data sources: glucose, physiological, activity data and nutrition logs.

\subsubsection{Data Synchronization and Alignment}
Data streams from CGM, wristband, and nutrition logs are first temporally aligned. Since Embrace Plus records data in UTC, all timestamps are converted to a common local timezone to ensure consistency with CGM and dietary logs. The CGM-derived signals are then upsampled to a uniform one-minute resolution to match the temporal granularity of wristband data. Missing values arising from sensor dropouts are handled using forward filling for short gaps ($\leq30$ minutes) and exclusion for longer discontinuities ($>30$ minutes).

\subsubsection{Meal Event Identification and Aggregation}
Meal events are extracted from nutrition logs. Due to the tendency of users to log multiple food items within a short time span, temporally adjacent meal entries occurring within a 30-minute interval are grouped into a single meal cluster. Each cluster is represented by its latest timestamp, corresponding to the final logged food item. The total macronutrient composition of the meal is computed as the sum of carbohydrates, protein and fat across all entries within the cluster.

\subsubsection{Feature Extraction}
For each event at time $t_m$, a feature vector $\mathbf{x}$ is constructed using data from a two-hour pre-meal window $[t_m - 2\text{h}, t_m)$. The extracted features include statistical summaries (mean and standard deviation) of wristband-derived signals, including step counts, activity counts, METs, EDA, skin temperature, and pulse rate. Subjects' age and BMI are also concatenated to $\mathbf{x}$ as stationary features.

Additionally, glycemic context features are computed from CGM data. Specifically, the pre-meal glucose level is defined as the last observed glucose value prior to $t_m$, and the short-term glucose trend is quantified as the slope of glucose values over the preceding 30-minute interval, estimated via linear regression. Both features capture the baseline glycemic state and recent dynamics prior to food intake.

\subsubsection{Target Variable Construction}
The forecast target $y$ was defined as the maximum glucose value observed within a two-hour postprandial window $(t_m, t_m + 2\text{h}]$. This definition aligns with clinical understanding of postprandial glucose excursions and captures peak glycemic response following a meal.

\subsubsection{Train/test split}
To prevent subject-level data leakage and ensure better generalization, the dataset is partitioned at subject level. Specifically, a 10/3 subject-wise split is followed, with ten participants being randomly selected for model training and the remaining three participants are set aside for evaluation and meal plan generation.

The data processing pipeline leaves us with \textbf{498} factual samples. Train/test split leaves \textbf{376} for training and \textbf{122} for test. Two samples are shown in Table~\ref{sample} as examples. Of the 17 features, we consider \textit{Carb}, \textit{Protein}, and \textit{Fat} as modifiable for meal recommendation.

\subsubsection{Supplementing the training data}

Given the limited size of the train dataset ($376$ samples), the training data is supplemented with an additional dataset from the MealMeter \cite{Arefeen2025MealMeterUM} project (IRB \#15102) comprising $12$ subjects ($n=168$ samples). In this cohort, participants were monitored in a controlled lab setting while consuming standardized meals and wearing a Dexcom G6 Pro and an Empatica E4 wrist-worn device.

The integration of these datasets introduces a modality mismatch due to differences in sensing platforms. The original dataset, collected using the Empatica Embrace Plus, provides device-derived features like step counts, METs, and activity counts by default. In contrast, the Empatica E4 provides raw high-frequency tri-axial accelerometry without such pre-computed features. Since several activity-related features are not directly available in the supplementary dataset, they are reconstructed from accelerometry using established methods. Physical activity intensity is estimated using Euclidean Norm Minus One (ENMO), computed from tri-axial acceleration and expressed in milli-gravitational units (mg), which has been widely adopted as a proxy for movement intensity in wrist-worn accelerometry \cite{vanHees2013SeparatingMA, Hildebrand2014AgeGC}. Step counts are estimated using a peak-detection approach applied to band-pass filtered acceleration (0.5--3 Hz), with additional cadence-based constraints to retain physiologically plausible walking periods. This approach is consistent with open-source wrist accelerometry frameworks like GGIR and related implementations \cite{vanHees2011EstimationOD}.

In contrast, MET cannot be reliably estimated from wrist accelerometry alone without population-specific calibration and additional physiological signals. Prior work has demonstrated substantial variability in energy expenditure estimation from wrist-worn devices across activity types and individuals \cite{Strath2013GuideTT, Migueles2017AccelerometerDC}. Therefore, MET is not reconstructed and is instead omitted entirely from training data to avoid introducing biased or unreliable estimates.

With instances from MealMeter, the training data has \textbf{544} samples.

\section{Experimental Setup}

\subsection{Model}
The forecasting task is formulated as a supervised regression problem. To evaluate the effectiveness of the proposed feature representation, we considered a set of eight regression models spanning linear, ensemble, and gradient-boosting paradigms: Ridge regression, Elastic Net, Random Forest, Extra Trees, Gradient Boosting, Histogram-based Gradient Boosting, XGBoost, and LightGBM.

For robust and reproducible model selection, hyperparameters for all models were optimized using randomized search over predefined distributions. Specifically, 3-fold cross-validation strategy is followed with shuffled splits to evaluate candidate configurations, using root mean squared error (RMSE) as the primary selection criterion.

After hyperparameter tuning, each model is retrained using the optimal configuration on the full training dataset and evaluated on the held-out test set.

\subsection{Counterfactual Optimizer}
Building on the aforementioned CF formulation, CF optimization is implemented using differential evolution, a population-based global optimizer. CFs are generated independently for all eight predictive models by optimizing the modifiable macronutrient variables, while keeping the remaining pre-meal context fixed. To improve robustness, instances whose original predicted glucose response was already below $140$ mg/dL are skipped and not used for CF generation. The search space is constrained to physiologically plausible values using $\epsilon = 10^{-6}$ as the lower bound for all macronutrients. Differential evolution is run with a population size of $15$ and a maximum of $80$ iterations. 

\subsection{LLMs}
Some of the latest LLMs have been used to map the CF generated macronutrient constraints into human understandable food items-
\begin{itemize}
    \item gemma-4-26b-a4b-it or Gemma 4 26B (April, 2026)
    \item GPT-5.4 (March, 2026)
    \item GPT-5.4-mini (March, 2026)
    \item llama-3.3-70b-versatile or Llama 3.3 70B (December, 2024)
    \item moonshotai-kimi-k2-instruct-0905 or Kimi K2 Instruct (September, 2025)
\end{itemize}

\subsection{Validation Metrics}
Different set of validation metrics have been used for evaluating different components of the MetaPlate pipeline.
\subsubsection{Regression model} The forecasting model is evaluated using a set of standard regression performance metrics: RMSE, MAE, median absolute error (MedAE), coefficient of determination ($R^2$), explained variance (Exp. Var.), Pearson correlation coefficient ($r$), MAPE, and sMAPE.
\subsubsection{Counterfactuals} CFs are validated using the metrics below-

\textbf{\textit{Validity}} assesses whether the generated CFs achieve the desired glycemic outcome under a regression setting. A CF is considered valid if the predicted postprandial glucose level falls below a predefined evaluation threshold $\tau_{\text{eval}} = 140$ mg/dL.

\begin{equation}
\textit{validity} = \frac{1}{|\mathcal{X}|} \sum_{(\mathbf{x}, \mathbf{m}_0) \in \mathcal{X}} 
\mathbbm{1}\big(f_\theta(\mathbf{x}, \mathbf{m}^*) \leq \tau_{\text{eval}}\big)
\end{equation}

where, $\mathcal{X}$ is the set of factual instances 

\textbf{\textit{Distance}} measures how close a counterfactual meal $\mathbf{m}^*$ is to the original meal $\mathbf{m}_0$, computed over the actionable feature set $\mathcal{A}$. Lower distance indicates smaller and more realistic modifications.

We compute both $L_1$ and $L_2$ distances:

\begin{equation}
L_1(\mathbf{m}_0, \mathbf{m}^*) = \sum_{j \in \mathcal{A}} |m_j^* - m_{0,j}|
\end{equation}

\begin{equation}
L_2(\mathbf{m}_0, \mathbf{m}^*) = \sqrt{\sum_{j \in \mathcal{A}} (m_j^* - m_{0,j})^2}
\end{equation}

To ensure comparability across methods, both distances are max-normalized across all valid counterfactuals:

\begin{equation}
\tilde{L}_k = \frac{L_k}{\max(L_k)}, \quad k \in \{1,2\}
\end{equation}

\textbf{\textit{Sparsity}} measures the number of actionable features that change between the original meal $\mathbf{m}_0$ and the counterfactual meal $\mathbf{m}^*$. Lower sparsity improves interpretability by minimizing the number of modifications required.

To avoid counting negligible variations, a tolerance of 1 gram is applied. A feature is considered changed only if:

\begin{equation}
|m_j^* - m_{0,j}| > 1, \quad j \in \mathcal{A}
\end{equation}

The sparsity is computed as:

\begin{equation}
\textit{sparsity} = \frac{1}{|CF_{\text{valid}}|} \sum_{\mathbf{m}^* \in CF_{\text{valid}}} 
\sum_{j \in \mathcal{A}} \mathbbm{1}\big(|m_j^* - m_{0,j}| > 1\big)
\end{equation}
\subsubsection{LLM mappings} 
The LLM-based meal mapping module is evaluated along three dimensions-

\textit{\textbf{Constraint Satisfaction (RMSE)}}
measures how closely the LLM generated meal matches the target macronutrient profile using root mean squared error (RMSE) for each macronutrient:

\begin{equation}
\mathrm{RMSE}_j = \sqrt{\frac{1}{N} \sum_{i=1}^{N} \left(m_{ij}^{LLM} - m_{ij}^{*}\right)^2}
\end{equation}

where $j \in \{C, P, F\}$ denotes carbohydrates, protein, and fat, respectively, $m_{ij}^{*}$ is the target amount for nutrient $j$ in the $i$-th meal, and $m_{ij}^{LLM}$ is the corresponding amount generated by the LLM. Lower RMSE indicates better adherence to the target macronutrient constraints.

\textit{\textbf{Glycemic Consistency}} assesses whether the generated meals preserve the intended glycemic outcome by re-evaluating them using the forecasting model. Specifically, we compute the proportion of generated meals that satisfy the target glucose constraint:

\begin{equation}
\textit{Consistency} = \frac{1}{N} \sum_{i=1}^{N} \mathbbm{1}\big(f_\theta(\mathbf{x}_i, \mathbf{m}^{LLM}_i) \leq \tau \big)
\end{equation}

where $N$ denotes the total number of generated meals, $\mathbbm{1}(\cdot)$ is the indicator function, $f_\theta(\mathbf{x}_i, \mathbf{m}^{LLM}_i)$ is the predicted postprandial glucose level for the $i$-th generated meal, and $\tau$ is the target glucose threshold.

\textit{\textbf{Food-Item Diversity}} quantifies diversity among the food items produced by an LLM. Each generated food item is embedded using a pretrained text embedding model and compute the average pairwise distance between all item embeddings. Let $\mathbf{e}_i$ denote the embedding of food item $i$, and let $N$ be the total number of generated items. Diversity is defined as:
\begin{equation}
\textit{Diversity} = \frac{2}{N(N-1)} \sum_{i=1}^{N-1} \sum_{j=i+1}^{N} d(\mathbf{e}_i, \mathbf{e}_j)
\end{equation}
where $d(\cdot,\cdot)$ is a Euclidean distance function in the embedding space. Higher values indicate greater diversity among the generated food items. Food-item embeddings are generated using a pretrained Sentence-BERT model (all-MiniLM-L6-v2) \cite{Wang2020MiniLMDS}.

\subsubsection{Expert-Based Validation of the Interventions}

We evaluate the clinical relevance and practical applicability of the generated meal interventions through expert assessment.

\begin{table*}[t!]
\centering
\caption{Regression performance on the test set. Lower is better for RMSE, MAE, MedAE, MAPE, and sMAPE; higher is better for $R^2$, Explained Variance, and Pearson $r$.}
\label{tab:final_model_results}
\small
\begin{tabular}{lC{1.15cm}C{1.1cm}C{1.3cm}C{0.8cm}C{1.5cm}C{1.55cm}C{1.75cm}C{1.9cm}}
\toprule
Model & RMSE $\downarrow$ & MAE $\downarrow$ & MedAE $\downarrow$ & $R^2$ $\uparrow$ & Exp. Var. $\uparrow$ & Pearson $r$ $\uparrow$ & MAPE (\%) $\downarrow$ & sMAPE (\%) $\downarrow$ \\
\midrule
ElasticNet           & \textbf{16.458} & 12.586 & \textbf{10.003} & \textbf{0.472} & \textbf{0.473} & 0.689 & 8.832 & 8.843 \\
ExtraTrees           & 16.573 & \textbf{12.552} & 10.234 & 0.464 & 0.465 & \textbf{0.691} & 8.951 & 8.848 \\
XGBoost              & 16.602 & 12.588 & 10.587 & 0.463 & 0.465 & 0.685 & \textbf{8.813} & \textbf{8.821} \\
Ridge                & 16.975 & 12.867 & 10.550 & 0.438 & 0.447 & 0.685 & 8.897 & 9.019 \\
GradientBoosting     & 17.532 & 13.406 & 11.515 & 0.401 & 0.401 & 0.651 & 9.526 & 9.448 \\
LightGBM             & 17.559 & 13.651 & 11.501 & 0.399 & 0.402 & 0.643 & 9.630 & 9.663 \\
RandomForest         & 17.721 & 13.673 & 11.698 & 0.388 & 0.388 & 0.631 & 9.781 & 9.631 \\
HistGradientBoosting & 17.896 & 13.571 & 11.151 & 0.375 & 0.379 & 0.615 & 9.545 & 9.525 \\
\bottomrule
\end{tabular}
\end{table*}

\textbf{\textit{Case-level Evaluation:}} 
Experts were provided with the subject context, predicted postprandial glucose response, and the corresponding MetaPlate-generated meal recommendation for each case. Each intervention was evaluated along four dimensions:

\begin{itemize}
    \item Glycemic appropriateness (ability to maintain postprandial glucose below target)
    \item Portion suitability
    \item Nutritional alignment with dietary guidelines
    \item Recommendation likelihood
\end{itemize}

Each dimension was rated on a 10-point Likert scale. Performance is summarized by averaging ratings across all cases and experts for each evaluation dimension. Specifically, for a given criterion $k$, the average score is computed as:

\begin{equation}
\bar{s}_k = \frac{1}{N M} \sum_{i=1}^{N} \sum_{j=1}^{M} s_{ijk}
\end{equation}

where $N$ denotes the total number of evaluated cases, $M$ denotes the number of experts, and $s_{ijk}$ represents the rating assigned by the $j$-th expert to the $k$-th criterion for the $i$-th case.

\textbf{\textit{System-level Evaluation:}}
In addition to case-level assessment, MetaPlate is also evaluated with established human-centered evaluation frameworks for explainable AI \cite{Holzinger2019MeasuringTQ}. Accordingly, experts evaluated the MetaPlate system across five aspects:

\begin{itemize}
    \item Ease of use
    \item Consistency with clinical knowledge
    \item Trustworthiness of recommendations
    \item Practical usability
    \item Interpretability of meal composition
\end{itemize}

\section{Results}

\subsection{Forecasting Performance}

Post-meal glucose forecasting performance across models is summarized in Table~\ref{tab:final_model_results}. The predictive accuracies are comparable. Therefore, the models are further assessed based on the quality of CFs generated during the CF optimization process using the models. As reported in Table~\ref{tab:minmax_model_comparison}, LightGBM achieves the lowest RMSE toward the target value of 140, the smallest $l_2$ distance, and the second-lowest $l_1$ distance among all models. Hence, LightGBM is used in all subsequent experiments.

\begin{table}[h]
\centering
\caption{Model comparison using normalized $l_1$, $l_2$ distance and RMSE to target 140. Lower scores are better}
\begin{tabular}{lccc}
\toprule
Model & RMSE to 140 $\downarrow$ & $l_1$ Dist. $\downarrow$ & $l_2$ Dist. $\downarrow$ \\
\midrule
ElasticNet           & 14.487 & 0.331 & 0.355 \\
ExtraTrees           & 11.537 & 0.225 & 0.266 \\
GradientBoosting     & 7.816  & 0.183 & 0.228 \\
HistGradientBoosting & 11.414 & \textbf{0.140} & 0.187 \\
LightGBM             & \textbf{7.030}  & 0.142 & \textbf{0.176} \\
RandomForest         & 9.807  & 0.166 & 0.210 \\
Ridge                & 7.067  & 0.352 & 0.401 \\
XGBoost              & 8.970  & 0.183 & 0.221 \\
\bottomrule
\end{tabular}
\label{tab:minmax_model_comparison}
\end{table}

\subsection{Counterfactual Performance}

Table~\ref{cf_method_comparison} presents a comparative evaluation of the proposed MetaPlate CF optimization framework against two widely used baseline methods, namely Wachter et al. and DiCE, across four key metrics: validity, normalized L1 and L2 distance and sparsity.

MetaPlate achieves the highest validity $0.660$, substantially outperforming both Wachter $0.540$ and DiCE $0.500$. This indicates that a larger proportion of generated CF meals successfully satisfy the target glycemic constraint ($\leq 140$ mg/dL). The improvement in validity can be attributed to the explicit addition of domain-specific constraints (e.g., limiting carbohydrate increase) and the use of a target-aware loss function that directly penalizes deviations from the glycemic threshold.

In terms of proximity to the original meal, MetaPlate also yields lower normalized $L_1$ $(0.230)$ and $L_2$ $(0.258)$ distances compared to Wachter $(0.326 \text{ and } 0.350)$ and DiCE $(0.419 \text{ and } 0.465)$. This means that MetaPlate produces more minimal, realistic modifications and preserves the structure of the original meal while achieving the desired outcome. In contrast, DiCE exhibits the largest distances, which indicates substantial deviations from the original meal composition and may reduce practical feasibility despite its ability to generate diverse CFs.

\begin{table}[b]
\centering
\caption{Comparison of CF methods using validity, normalized distance, and sparsity}
\begin{tabular}{lcccc}
\toprule
Method & \cellcolor{blue!18}Validity  $\uparrow$ & \cellcolor{gray!18}$L_1$ Dist.  $\downarrow$ & \cellcolor{red!25}$L_2$ Dist.  $\downarrow$ & \cellcolor{yellow!25}Sparsity  $\downarrow$ \\
\midrule
MetaPlate & \textbf{0.660} & \textbf{0.230} & \textbf{0.258} & 2.030 \\
Wachter   & 0.540 & 0.326 & 0.350 & 2.222 \\
DiCE      & 0.500 & 0.419 & 0.465 & \textbf{1.800} \\
\bottomrule
\end{tabular}
\label{cf_method_comparison}
\end{table}

\begin{table*}[h]
\centering
{\renewcommand{\arraystretch}{1.1}
\caption{LLM Performance Comparison Across Nutritional Accuracy, Glycemic Consistency, and Diversity}
\begin{tabular}{lccccccccc}
\toprule
\textbf{LLM} & \textbf{n} & \cellcolor{orange!12}\textbf{Carb RMSE $\downarrow$} & \cellcolor{green!15}\textbf{Prot RMSE $\downarrow$} & \cellcolor{red!10}\textbf{Fat RMSE $\downarrow$} & \cellcolor{lime!18}\textbf{Gly. Const.} $\uparrow$ & \cellcolor{blue!10}\textbf{Diversity} $\uparrow$ & \cellcolor{yellow!15}\textbf{\# Items} $\uparrow$ & \cellcolor{teal!15}\textbf{Unique items} $\uparrow$ \\
\midrule
Gemma 4 26B & 50 & \textbf{0.187} & \textbf{0.147} & \textbf{0.211} & \textbf{0.700} & \textbf{1.123} & \textbf{199} & 80 \\
GPT-5.4 & 50 & 4.215 & 2.753 & 1.729 & 0.320 & 1.040 & 192 & 37 \\
GPT-5.4-mini & 50 & 6.334 & 2.941 & 3.189 & 0.320 & 0.962 & 185 & 27 \\
Llama 3.3 70B & 50 & 11.168 & 3.216 & 5.012 & 0.440 & 0.952 & 171 & 26 \\
Kimi K2 Instruct & 50 & 8.596 & 8.580 & 6.652 & 0.520 & 1.105 & 195 & \textbf{121} \\
\bottomrule
\end{tabular}}
\label{llm_performance}
\end{table*}

DiCE achieves the lowest sparsity $(1.800)$ i.e., fewer features are modified on average. However, this comes at the cost of reduced validity and significantly larger perturbations in magnitude. MetaPlate exhibits a balanced trade-off between the number of macro-nutrient changes and their magnitude by maintaining a competitive sparsity level $(2.030)$, slightly higher than DiCE but lower than Wachter $(2.222)$. Importantly, the slightly higher sparsity in MetaPlate reflects deliberate adjustments across multiple macronutrients to ensure both physiological plausibility and adherence to glycemic constraints.

Our analysis shows that MetaPlate achieves a favorable balance between effectiveness and practicality, generating counterfactual meal recommendations that are more likely to meet glycemic targets while requiring smaller and more logical modifications compared to existing CF methods.

\subsection{LLM Performance}

The LLMs are assessed on the basis of macronutrient constraint satisfaction, glycemic consistency, and food-item diversity. The results are listed in Table~\ref{llm_performance}. 

\begin{figure}[!h]
\centerline{\includegraphics[width=\columnwidth]{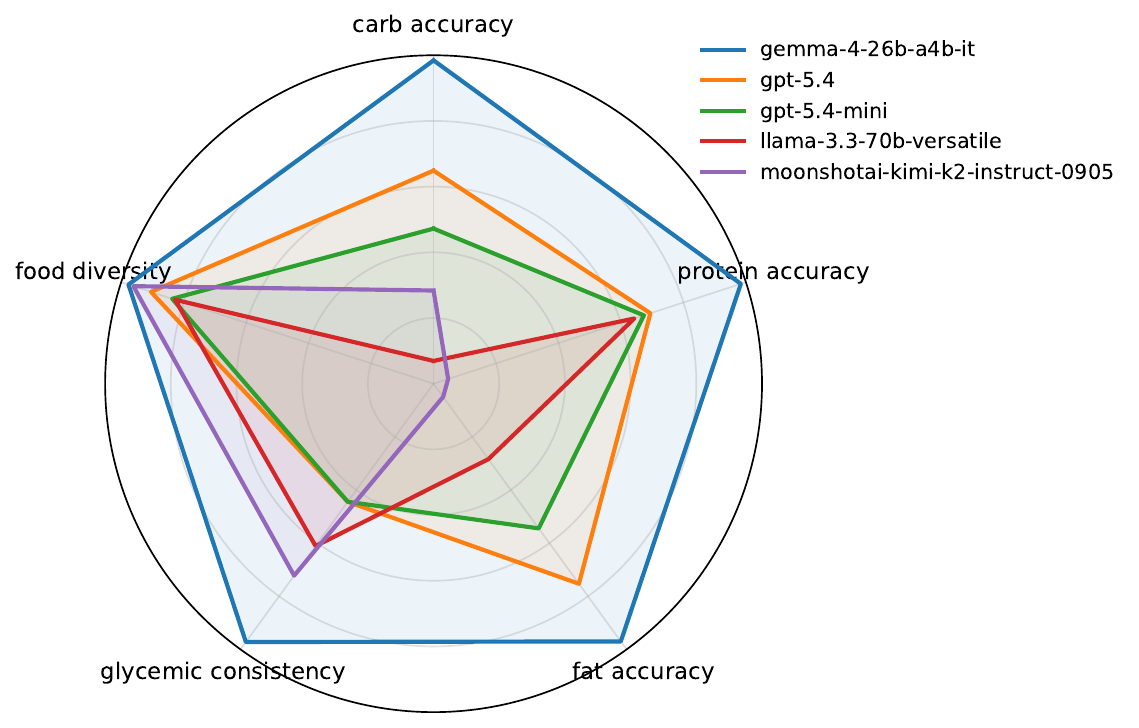}}
\caption{LLM comparison across RMSE, glycemic consistency, and diversity. For visualization purposes, RMSE-based metrics are normalized and inverted to accuracy such that higher values indicate better performance. Specifically, lower RMSE corresponds to higher normalized accuracy in the radar plot.}
\label{diversity}
\end{figure}

The radar plot in Figure~\ref{diversity} provides a normalized visualization of the same results, where the RMSE-based metrics are inverted such that higher values indicate better adherence to the target nutritional profile. 

Among the evaluated LLMs, \texttt{Gemma 4 26B} achieves the best overall nutritional fidelity, with the lowest carbohydrate, protein, and fat RMSE values $(0.187, 0.147, \text{ and } 0.211\text{, respectively})$ and the highest glycemic consistency $(0.700)$. This indicates that the LLM most accurately maps the optimized macronutrient targets to feasible meal suggestions while preserving the intended glycemic effect. The corresponding radar profile in Figure~\ref{diversity} is the most balanced and expansive across the accuracy, consistency and diversity axes, reflecting its superior alignment with the CF targets.

In contrast, \texttt{GPT-5.4}, \texttt{GPT-5.4-mini}, and \texttt{Llama 3.3 70B} exhibit substantially larger nutrient RMSE values, indicating weaker adherence to the prescribed macronutrient constraints. Although \texttt{Llama 3.3 70B} attains a moderate glycemic consistency of $0.440$, its nutrient-level errors remain comparatively high, suggesting that the generated meals are less tightly aligned with the optimized nutritional profile. The radar plot reflects this behavior through a narrower polygon on the nutrient accuracy dimensions.

\texttt{Kimi K2 Instruct} shows an intermediate profile. While its carbohydrate, protein, and fat RMSE values are larger than those of \texttt{Gemma 4 26B}, it attains the highest food-item diversity $(1.105)$ and the largest number of unique items used for diversity analysis $(121)$, indicating broader lexical and compositional variability in the generated recommendations. However, this diversity does not translate into the best macronutrient matching, which suggests a trade-off between recommendation variety and constraint fidelity. This pattern is also evident in Figure~\ref{diversity}, where the diversity axis is relatively strong despite weaker nutrient-accuracy axes.

Based on our analysis, \texttt{Gemma 4 26B} appears to be the most suitable LLM for the MetaPlate retrieval-and-generation layer as it provides the most reliable balance between nutritional accuracy and glycemic consistency. At the same time, the diversity observed in \texttt{Kimi K2 Instruct} suggests that models with broader generative variability may be useful when recommendation richness is prioritized, albeit at the expense of tighter adherence to the target macronutrient constraints.

\subsection{Quality of the Interventions}

\begin{figure}[!h]
\centerline{\includegraphics[width=\columnwidth]{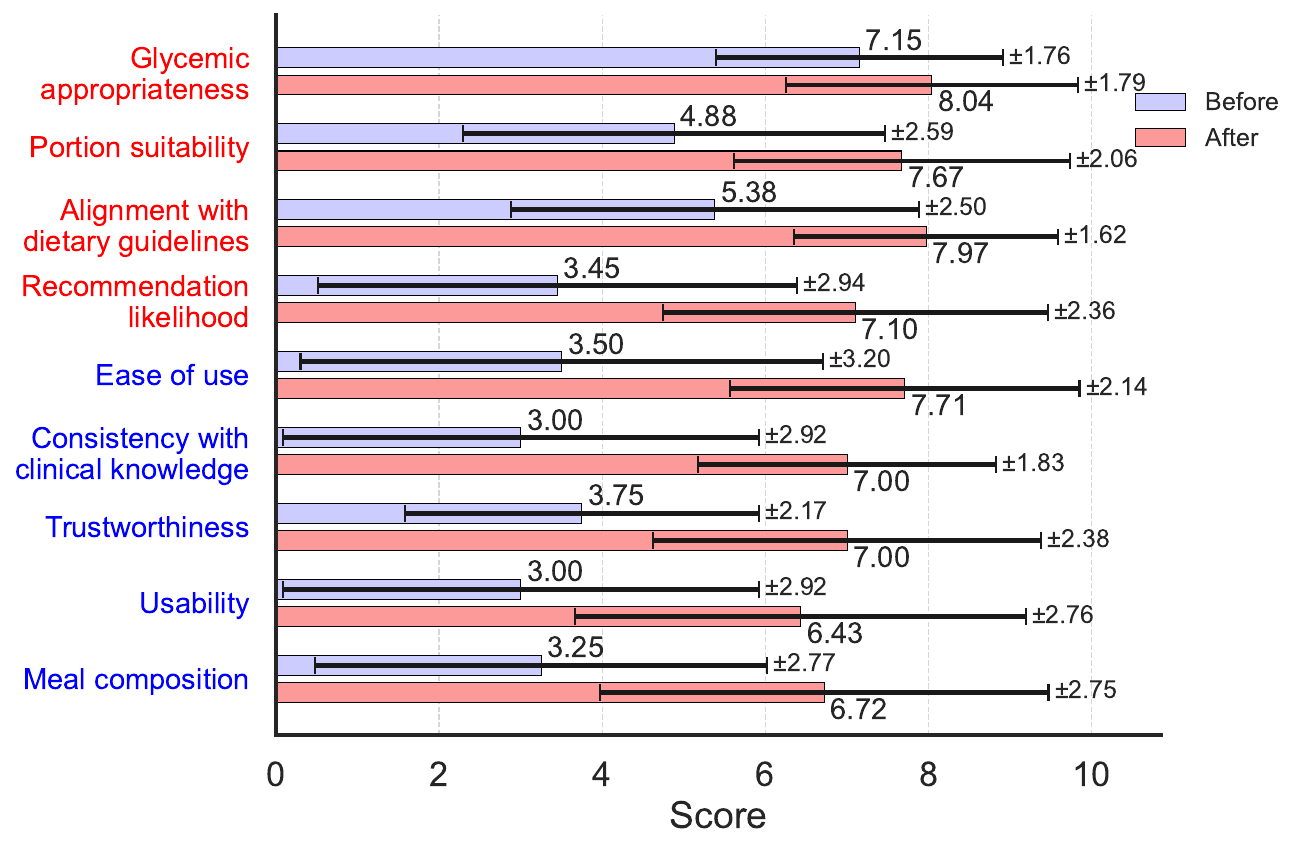}}
\caption{Comparison of expert evaluation scores before and after prompt refinement across case-level (red) and system-level (blue) dimensions. Ratings are reported on a 10-point Likert scale with error bars indicating standard deviation across experts and cases. Substantial improvements are observed across all dimensions following prompt redesign, particularly in portion suitability, recommendation likelihood, ease of use, and trustworthiness. Impact of expert-in-the-loop refinement are also noticeable on the realism, usability, and clinical alignment of MetaPlate-generated meal recommendations.}
\label{bar_plot}
\end{figure}

We evaluate the quality of MetaPlate-generated meal recommendations through a two-stage expert assessment involving registered dietitians. In the first round, four experts evaluated the meals generated using an initial prompt formulation. Based on their feedback, the prompt was redesigned by the team and the new meals are re-evaluated in a second round with six experts. All evaluators are registered dietitians with substantial clinical experience.

Figure~\ref{bar_plot} presents a comparison of expert ratings before and after prompt refinement across both case-level and system-level evaluation dimensions. At the case level, consistent improvements are observed across all four criteria. Glycemic appropriateness improves modestly $(7.15 \text{ to } 8.04$), indicating that even the initial formulation was reasonably aligned with glycemic goals. However, substantially larger gains are observed in portion suitability $(4.88 \text{ to } 7.67)$, nutritional alignment $(5.38 \text{ to } 7.97)$, and recommendation likelihood $(3.45 \text{ to } 7.10)$. The trends suggest that the primary limitations of the initial system were not in glycemic correctness, but in producing realistic, nutritionally coherent, and practically acceptable meals.

System-level evaluation further reinforces this trend. Metrics like ease of use $(3.50 \text{ to } 7.71)$, consistency with clinical knowledge $(3.00 \text{ to } 7.00)$, trustworthiness $(3.75 \text{ to } 7.00)$, usability $(3.00 \text{ to } 6.43)$, and meal composition quality $(3.25 \text{ to } 7.72)$ all show substantial improvement after refinement. Notably, the largest gains are observed in ease of use and perceived clinical validity. This means that the resulting meals from the redesigned system is significantly more aligned with expert expectations for real-life applicability.

The improvements can be attributed to changes made during prompt redesign. The initial prompt emphasized strict macronutrient matching, which often resulted in unrealistic or fragmented meal compositions, including snack-like outputs, excessive reliance on filler ingredients (e.g., nuts, almonds), and implausible portion sizes. In contrast, the redesigned prompt explicitly enforces meal structure (protein + carbohydrate + produce), realistic portioning, and dietary coherence, while slightly relaxing strict macro constraints when necessary to preserve plausibility. Additionally, it prioritizes clinical realism and balanced meal composition over exact numerical matching, and introduces explicit checks for meal validity.

The results show that expert-in-the-loop prompt refinement significantly enhances the quality, interpretability, and practical usability of generated interventions. This is consistent with the fact that co-creation and integration of domain knowledge and human feedback is necessary when deploying LLM-based systems for clinically relevant decision support.

In addition to quantitative improvements, qualitative analysis of expert feedback reveals a clear shift in the nature and severity of system limitations between the two evaluation rounds. In the initial evaluation, feedback was dominated by fundamental structural issues, including non-meal-like outputs (e.g., “this is a snack, not a meal”), unrealistic food combinations (e.g., “meat and almonds only”), and extreme macronutrient modifications that were deemed clinically impractical (e.g., excessive carbohydrate restriction). Experts frequently questioned the plausibility, palatability, and clinical applicability of the recommendations and indicated that the system outputs were often not usable in real-world settings.

Following prompt refinement, these concerns were substantially reduced. Feedback from the second round shifted toward fine-grained adjustments, like modest portion tuning, minor macronutrient balancing, and flavor enhancements. Experts more frequently described the recommendations as “reasonable”, “balanced”, and “clinically appropriate”, with suggestions focusing on improving rather than rejecting the outputs. This transition from rejection-level feedback to refinement-level feedback indicates that the redesigned system successfully addresses core structural and plausibility issues, resulting in recommendations that are both interpretable and actionable in practice.

These qualitative observations align with the quantitative improvements reported in Figure~\ref{bar_plot}, where the largest gains are observed in \textit{portion suitability}, \textit{nutritional alignment}, and \textit{recommendation likelihood} — dimensions directly related to meal realism and practical usability.

We categorize the primary failure modes of the initial system into three groups: (1) structural issues, including generation of snack-like or incomplete meals; (2) quantitative inconsistencies, like unrealistic macronutrient distributions and portion sizes; and (3) practical usability concerns, including lack of dietary coherence, variety, and palatability. These failure modes are substantially mitigated after prompt refinement.

Notably, the absence of repeated concerns such as over-reliance on almonds and lack of meal structure in the second round suggests that including explicit meal composition constraints and diversity controls effectively mitigates common failure modes of LLM-based food generation.

\section{Discussion}

\textbf{Motivation for LLM-RAG-Based Food Mapping:}
Mapping user-provided meal descriptions to concrete food items in a structured database (e.g., USDA FoodData Central) is inherently challenging due to the noisy, heterogeneous, and context-dependent nature of food text. Prior work has shown that naive or straightforward lexical matching approaches struggle to accurately identify specific food items from such databases, as superficial word overlap can lead to incorrect matches and misclassification in the presence of brand names, preparation details, and ambiguous terminology \cite{Metwally2021LearningPF}. In the context of MetaPlate, a purely lexical or rule-based retrieval approach would be insufficient as it cannot reason over combinations of foods, provide healthy food suggestions, enforce dietary coherence, or ensure that aggregated nutrient profiles align with the target constraints. To overcome this challenges, MetaPlate uses a LLM-based RAG module that enables semantic matching and compositional reasoning over candidate foods.

\textbf{Implications for clinical usability and adoption:}
Expert feedback also highlights an important consideration for clinical adoption: systems that recommend entirely new meals may face resistance if they do not align with users’ existing dietary habits. Instead, approaches that preserve familiar food choices while suggesting realistic modifications are more consistent with human adaptability based on dietary counseling practices.

\textbf{Design considerations for LLM-based dietary systems:}
Insights from experts illustrates a broader limitation of LLM-based generation systems in healthcare applications, where unconstrained optimization may produce technically valid but clinically implausible outputs. Integrating domain-specific constraints, structured reasoning, and expert feedback is therefore prerequisite to ensure both accuracy and usability before deployment.

\section{Limitations and Future Plan}
\textbf{Limited dataset size:} 
The current study is based on a relatively small cohort collected under an ongoing data acquisition effort, which may limit the generalizability and robustness of the learned models. Although we partially mitigate this constraint through data augmentation using an external dataset, the overall sample size remains modest compared to large-scale studies. As a result, the predictive and counterfactual components may not fully capture the variability in glycemic responses across diverse populations and contexts. In future work, we plan to expand the cohort by recruiting additional participants and collecting longer-duration recordings under free-living conditions. This will unlock the development of more robust models, including deep learning architectures capable of capturing complex temporal and multimodal dependencies, and support more comprehensive and statistically rigorous evaluations of the proposed framework.

\textbf{Evaluation in clinical trials:} 
While MetaPlate demonstrates promising performance in offline evaluation and expert-based assessments, it has not yet been validated in an active intervention setting. As such, its practical effectiveness, user adherence, and feasibility under real-life constraints, like dietary preferences, accessibility of food items, and behavioral factors—remain to be established. The current evaluation does not capture longitudinal outcomes or the impact of repeated system use on glycemic control. To address this limitation, future work will involve deploying MetaPlate as a smartphone-based application and conducting a prospective clinical trial. Specifically, we plan to collect an initial baseline period of four weeks without intervention, followed by a four-week intervention phase during which participants will actively use MetaPlate for meal decision support. Assessment will include real-world usability, adherence, and effectiveness in maintaining in-range postprandial glucose levels.

\textbf{LLM-generated meal inconsistencies:} 
Despite prompt refinement and constraint-based generation, the LLM occasionally produces meal suggestions that deviate from intended guidelines, like including restricted items (e.g., nuts or almonds) or generating snack-like combinations instead of structured meals. These inconsistencies highlight limitations in controllability and faithfulness of current LLM outputs, particularly when translating numerical macronutrient targets into discrete food items. Although rare, discrepancies may arise between the generated meal composition and the corresponding nutritional values due to hallucination or imperfect grounding in the underlying food database. To address this limitation, future work will incorporate a post-generation verification and filtering module that cross-checks generated meals against the USDA FoodData Central database. This step will ensure that selected food items and portion sizes accurately reflect the target macronutrient composition and adhere to predefined dietary constraints, thereby improving reliability and reducing hallucination in the recommendation pipeline.

\textbf{Food availability and practicality constraints:} 
The current formulation of MetaPlate assumes access to the recommended food items, which may not always hold in real scenarios due to availability, cost, or user preferences. As a result, some generated meal suggestions may be impractical or difficult to adopt at the time of decision-making. Furthermore, the system currently focuses on generating alternative meals rather than adapting to what the user already has available. To address this limitation, future work will introduce mechanisms for availability-aware recommendation, where users can specify accessible food items or constraints. Additionally, the system can be extended to support modification-based recommendations by adjusting components of the user’s intended meal (e.g., reducing portion sizes or substituting specific ingredients). For instance, if a user plans to consume a meal such as chicken with rice, MetaPlate could suggest reducing the rice portion or rebalancing macronutrients to achieve the desired glycemic outcome. Enhancements like these would improve the practicality and applicability of the system.

\textbf{Device-specific generalizability:} 
MetaPlate is currently developed using data from specific devices (Dexcom G6 Pro and Empatica Embrace Plus), which may limit its generalizability across different sensing platforms. We plan to extend the framework to widely available consumer devices like Dexcom G7, Abbott FreeStyle Libre 3, Apple Watch and Samsung Galaxy Watch, which offer comparable sensing modalities. This transition is expected to be feasible given the overlap in sensor capabilities and would improve scalability and real-life adoption.

\section{Conclusion}
This paper presents MetaPlate, a counterfactual-guided, context-aware decision-support framework for personalized meal recommendation to mitigate postprandial hyperglycemia in healthy adults. MetaPlate extends beyond prediction to provide actionable dietary interventions. The framework formulates meal recommendation as a constrained counterfactual optimization problem and translates optimized macronutrient targets into human-understandable meal suggestions through an LLM-based RAG layer grounded in the USDA FoodData Central database. Experimental results demonstrate that MetaPlate generates valid and minimally modified counterfactual meals that satisfy glycemic constraints, while expert evaluation by registered dietitians confirms the clinical appropriateness, nutritional validity, and practical usability of the recommendations. Furthermore, expert-in-the-loop prompt refinement significantly enhances the realism, coherence, and trustworthiness of generated meals, underscoring the importance of incorporating domain expertise in LLM-driven health systems. MetaPlate holds potential as a scalable and interpretable tool for real-time, personalized dietary guidance and early prevention of glycemic dysregulation, with future work directed toward clinical validation and deployment in real-world settings.

\section*{Acknowledgment}
This work was supported in part by the National Institute of Diabetes and Digestive and Kidney Diseases of the National Institutes of Health under grant T32DK137525 and the National Science Foundation under grant IIS-2402650. Any opinions, findings, conclusions, or recommendations expressed in this material are those of the authors and do not necessarily reflect the views of the funding organization.



\begin{thebibliography}{10}
\providecommand{\url}[1]{#1}
\csname url@samestyle\endcsname
\providecommand{\newblock}{\relax}
\providecommand{\bibinfo}[2]{#2}
\providecommand{\BIBentrySTDinterwordspacing}{\spaceskip=0pt\relax}
\providecommand{\BIBentryALTinterwordstretchfactor}{4}
\providecommand{\BIBentryALTinterwordspacing}{\spaceskip=\fontdimen2\font plus
\BIBentryALTinterwordstretchfactor\fontdimen3\font minus \fontdimen4\font\relax}
\providecommand{\BIBforeignlanguage}[2]{{%
\expandafter\ifx\csname l@#1\endcsname\relax
\typeout{** WARNING: IEEEtran.bst: No hyphenation pattern has been}%
\typeout{** loaded for the language `#1'. Using the pattern for}%
\typeout{** the default language instead.}%
\else
\language=\csname l@#1\endcsname
\fi
#2}}
\providecommand{\BIBdecl}{\relax}
\BIBdecl

\bibitem{Jarvis2023ContinuousGM}
P.~R.~E. Jarvis, J.~L. Cardin, P.~M. Nisevich-Bede, and J.~P. McCarter, ``Continuous glucose monitoring in a healthy population: understanding the post-prandial glycemic response in individuals without diabetes mellitus.'' \emph{Metabolism: clinical and experimental}, p. 155640, 2023.

\bibitem{Giri2018ChronicHM}
B.~Giri, S.~Dey, T.~Das, M.~Sarkar, J.~Banerjee, and S.~K. Dash, ``Chronic hyperglycemia mediated physiological alteration and metabolic distortion leads to organ dysfunction, infection, cancer progression and other pathophysiological consequences: An update on glucose toxicity.'' \emph{Biomedicine \& pharmacotherapy = Biomedecine \& pharmacotherapie}, vol. 107, pp. 306--328, 2018.

\bibitem{Levitan2004IsNH}
E.~B. Levitan, Y.~Song, E.~S. Ford, and S.~Liu, ``Is nondiabetic hyperglycemia a risk factor for cardiovascular disease? a meta-analysis of prospective studies.'' \emph{Archives of internal medicine}, vol. 164 19, pp. 2147--55, 2004.

\bibitem{Phillips2023UncoveringPG}
N.~E. Phillips, T.-H. Collet, and F.~Naef, ``Uncovering personalized glucose responses and circadian rhythms from multiple wearable biosensors with bayesian dynamical modeling,'' \emph{Cell Reports Methods}, vol.~3, 2023.

\bibitem{Zeevi2015PersonalizedNB}
D.~A. Zeevi, T.~Korem, N.~Zmora, D.~Israeli, D.~Rothschild, A.~Weinberger, O.~Ben-Yacov, D.~Lador, T.~Avnit-Sagi, M.~Lotan-Pompan, J.~Suez, J.~A. Mahdi, E.~Matot, G.~Malka, N.~Kosower, M.~Rein, G.~Zilberman-Schapira, L.~Dohnalov{\'a}, M.~Pevsner-Fischer, R.~Bikovsky, Z.~Halpern, E.~Elinav, and E.~Segal, ``Personalized nutrition by prediction of glycemic responses.'' \emph{Cell}, vol. 163 5, pp. 1079--1094, 2015.

\bibitem{vanDoorn2021MachineLG}
W.~P. van Doorn, Y.~D. Foreman, N.~C. Schaper, H.~H. Savelberg, A.~Koster, C.~J.~H. van~der Kallen, A.~Wesselius, M.~T. Schram, R.~M.~A. Henry, P.~C. Dagnelie, B.~E. de~Galan, O.~Bekers, C.~D.~A. Stehouwer, S.~J.~R. Meex, and M.~C. Brouwers, ``Machine learning-based glucose prediction with use of continuous glucose and physical activity monitoring data: The maastricht study,'' \emph{PLoS ONE}, vol.~16, 2021.

\bibitem{Farahmand2025AttenGlucoMT}
E.~Farahmand, R.~R. Azghan, N.~T. Chatrudi, E.~Kim, G.~K. Gudur, E.~Thomaz, G.~Pedrielli, P.~K. Turaga, and H.~Ghasemzadeh, ``Attengluco: Multimodal transformer-based blood glucose forecasting on ai-readi dataset,'' \emph{2025 47th Annual International Conference of the IEEE Engineering in Medicine and Biology Society (EMBC)}, pp. 1--7, 2025.

\bibitem{Machiraju2025TimeAwareCF}
A.~Machiraju, E.~Farahmand, S.~B. Soumma, A.~Arefeen, C.~Johnston, and H.~Ghasemzadeh, ``Time-aware cross-attention for multi-modal sensor-based blood glucose forecasting,'' \emph{2025 IEEE 21st International Conference on Body Sensor Networks (BSN)}, pp. 1--4, 2025.

\bibitem{Zhu2024PopulationSpecificGP}
T.~Zhu, L.~Kuang, C.~Piao, J.~Zeng, K.~Li, and P.~Georgiou, ``Population-specific glucose prediction in diabetes care with transformer-based deep learning on the edge,'' \emph{IEEE Transactions on Biomedical Circuits and Systems}, vol.~18, pp. 236--246, 2024.

\bibitem{Shroff2023GlucoseAssistPB}
P.~Shroff, A.~Arefeen, and H.~Ghasemzadeh, ``Glucoseassist: Personalized blood glucose level predictions and early dysglycemia detection,'' \emph{2023 IEEE 19th International Conference on Body Sensor Networks (BSN)}, pp. 1--4, 2023.

\bibitem{Khamesian2025GlycemicAwareAA}
S.~Khamesian, A.~Arefeen, M.~A. Grando, B.~Thompson, and H.~Ghasemzadeh, ``Glycemic-aware and architecture-agnostic training framework for blood glucose forecasting in type 1 diabetes,'' 2025.

\bibitem{Soumma2026GlyRAGCR}
S.~B. Soumma and H.~Ghasemzadeh, ``Glyrag: Context-aware retrieval-augmented framework for blood glucose forecasting,'' \emph{ArXiv}, vol. abs/2601.05353, 2026.

\bibitem{Kalpakoglou2025AnAN}
K.~Kalpakoglou, L.~Calder{\'o}n-P{\'e}rez, N.~Boqu{\'e}, M.~Guldas, Çağla Erdoğan~Demir, L.~P. Gymnopoulos, and K.~Dimitropoulos, ``An ai-based nutrition recommendation system: technical validation with insights from mediterranean cuisine,'' \emph{Frontiers in Nutrition}, vol.~12, 2025.

\bibitem{Arefeen2022ComputationalFF}
A.~Arefeen, N.~Jaribi, B.~J. Mortazavi, and H.~Ghasemzadeh, ``Computational framework for sequential diet recommendation: Integrating linear optimization and clinical domain knowledge,'' \emph{2022 IEEE/ACM Conference on Connected Health: Applications, Systems and Engineering Technologies (CHASE)}, pp. 91--98, 2022.

\bibitem{Gavai2025AIdrivenPN}
A.~K. Gavai and J.~van Hillegersberg, ``Ai-driven personalized nutrition: Rag-based digital health solution for obesity and type 2 diabetes,'' \emph{PLOS Digital Health}, vol.~4, 2025.

\bibitem{Khamesian2025NutriGenPM}
S.~Khamesian, A.~Arefeen, S.~M. Carpenter, and H.~Ghasemzadeh, ``Nutrigen: Personalized meal plan generator leveraging large language models to enhance dietary and nutritional adherence,'' \emph{2025 47th Annual International Conference of the IEEE Engineering in Medicine and Biology Society (EMBC)}, pp. 1--7, 2025.

\bibitem{YANG2024100465}
Z.~Yang, E.~Khatibi, N.~Nagesh, M.~Abbasian, I.~Azimi, R.~Jain, and A.~M. Rahmani, ``Chatdiet: Empowering personalized nutrition-oriented food recommender chatbots through an llm-augmented framework,'' \emph{Smart Health}, vol.~32, p. 100465, 2024.

\bibitem{Zhang2024MOPIHFRSAM}
Z.~Zhang, Z.~Wang, T.~Ma, V.~S. Taneja, S.~Nelson, N.~H.~L. Le, K.~Murugesan, M.~Ju, N.~V. Chawla, C.~Zhang, and Y.~Ye, ``Mopi-hfrs: A multi-objective personalized health-aware food recommendation system with llm-enhanced interpretation,'' \emph{Proceedings of the 31st ACM SIGKDD Conference on Knowledge Discovery and Data Mining V.1}, 2024.

\bibitem{Crupi2021CounterfactualEA}
R.~Crupi, A.~Castelnovo, D.~Regoli, and B.~S.~M. Gonzalez, ``Counterfactual explanations as interventions in latent space,'' \emph{Data Mining and Knowledge Discovery}, vol.~38, pp. 2733 -- 2769, 2021.

\bibitem{Arefeen2023DesigningUB}
A.~Arefeen and H.~Ghasemzadeh, ``Designing user-centric behavioral interventions to prevent dysglycemia with novel counterfactual explanations,'' \emph{ArXiv}, vol. abs/2310.01684, 2023.

\bibitem{Arefeen2025GlyTwinDT}
A.~Arefeen, S.~Khamesian, M.~A. Grando, B.~Thompson, and H.~Ghasemzadeh, ``Glytwin: Digital twin for glucose control in type 1 diabetes through optimal behavioral modifications using patient-centric counterfactuals,'' \emph{ArXiv}, vol. abs/2504.09846, 2025.

\bibitem{Arefeen2024GlyManGM}
------, ``Glyman: Glycemic management using patient-centric counterfactuals,'' \emph{2024 IEEE EMBS International Conference on Biomedical and Health Informatics (BHI)}, pp. 1--5, 2024.

\bibitem{Barbiero2021EntropybasedLE}
P.~Barbiero, G.~Ciravegna, F.~Giannini, P.~Li'o, M.~Gori, and S.~Melacci, ``Entropy-based logic explanations of neural networks,'' \emph{ArXiv}, vol. abs/2106.06804, 2021.

\bibitem{mothilal2020dice}
R.~K. Mothilal, A.~Sharma, and C.~Tan, ``Explaining machine learning classifiers through diverse counterfactual explanations,'' in \emph{Proceedings of the 2020 Conference on Fairness, Accountability, and Transparency}, 2020, pp. 607--617.

\bibitem{Brughmans2021NICEAA}
D.~Brughmans and D.~Martens, ``Nice: an algorithm for nearest instance counterfactual explanations,'' \emph{Data Mining and Knowledge Discovery}, pp. 1--39, 2021.

\bibitem{DEOLIVEIRA2023}
R.~M.~B. {de Oliveira}, K.~Sörensen, and D.~Martens, ``A model-agnostic and data-independent tabu search algorithm to generate counterfactuals for tabular, image, and text data,'' \emph{European Journal of Operational Research}, 2023.

\bibitem{Arefeen2025MealMeterUM}
A.~Arefeen, S.~N. Fessler, S.~M. Mostafavi, C.~Johnston, and H.~Ghasemzadeh, ``Mealmeter: Using multimodal sensing and machine learning for automatically estimating nutrition intake,'' \emph{2025 47th Annual International Conference of the IEEE Engineering in Medicine and Biology Society (EMBC)}, pp. 1--6, 2025.

\bibitem{vanHees2013SeparatingMA}
V.~T. van Hees, L.~Gorzelniak, E.~C.~D. Le{\'o}n, M.~Eder, M.~R. Pias, S.~Taherian, U.~Ekelund, F.~Renstr{\"o}m, P.~W. Franks, A.~Horsch, and S.~Brage, ``Separating movement and gravity components in an acceleration signal and implications for the assessment of human daily physical activity,'' \emph{PLoS ONE}, vol.~8, 2013.

\bibitem{Hildebrand2014AgeGC}
M.~Hildebrand, V.~T. van Hees, B.~H. Hansen, and U.~Ekelund, ``Age group comparability of raw accelerometer output from wrist- and hip-worn monitors.'' \emph{Medicine and science in sports and exercise}, vol. 46 9, pp. 1816--24, 2014.

\bibitem{vanHees2011EstimationOD}
V.~T. van Hees, F.~Renstr{\"o}m, A.~Wright, A.~Gradmark, M.~Catt, K.~Y. Chen, M.~L{\"o}f, L.~J.~C. Bluck, J.~Pomeroy, N.~J. Wareham, U.~Ekelund, S.~Brage, and P.~W. Franks, ``Estimation of daily energy expenditure in pregnant and non-pregnant women using a wrist-worn tri-axial accelerometer,'' \emph{PLoS ONE}, vol.~6, 2011.

\bibitem{Strath2013GuideTT}
S.~J. Strath, L.~A. Kaminsky, B.~E. Ainsworth, U.~Ekelund, P.~S. Freedson, R.~A. Gary, C.~R. Richardson, D.~T. Smith, and A.~M. Swartz, ``Guide to the assessment of physical activity: Clinical and research applications: a scientific statement from the american heart association.'' \emph{Circulation}, vol. 128 20, pp. 2259--79, 2013.

\bibitem{Migueles2017AccelerometerDC}
J.~H. Migueles, C.~Cadenas-S{\'a}nchez, U.~Ekelund, C.~D. Nystr{\"o}m, J.~Mora-Gonzalez, M.~L{\"o}f, I.~Labayen, J.~R. Ruiz, and F.~B. Ortega, ``Accelerometer data collection and processing criteria to assess physical activity and other outcomes: A systematic review and practical considerations,'' \emph{Sports Medicine}, vol.~47, pp. 1821--1845, 2017.

\bibitem{Wang2020MiniLMDS}
W.~Wang, F.~Wei, L.~Dong, H.~Bao, N.~Yang, and M.~Zhou, ``Minilm: Deep self-attention distillation for task-agnostic compression of pre-trained transformers,'' \emph{ArXiv}, vol. abs/2002.10957, 2020.

\bibitem{Holzinger2019MeasuringTQ}
A.~Holzinger, A.~M. Carrington, and H.~M{\"u}ller, ``Measuring the quality of explanations: The system causability scale (scs),'' \emph{Kunstliche Intelligenz}, vol.~34, pp. 193 -- 198, 2019.

\bibitem{Metwally2021LearningPF}
A.~A. Metwally, A.~K. Leong, A.~Desai, A.~Nagarjuna, D.~Perelman, and M.~P. Snyder, ``Learning personal food preferences via food logs embedding,'' \emph{2021 IEEE International Conference on Bioinformatics and Biomedicine (BIBM)}, pp. 2281--2286, 2021.

\end{thebibliography}
\end{document}


\title{Supplementary Materials for MetaPlate}
\maketitle

\section{Model Fine-Tuning Procedure}

Hyperparameter optimization was conducted using randomized search. For linear and tree-based models (Ridge, Elastic Net, Random Forest, Extra Trees, Gradient Boosting, HistGradientBoosting, and XGBoost), we used \textit{RandomizedSearchCV} with 3-fold cross-validation (KFold, $n=3$, shuffle=True, random state=42), optimizing for negative RMSE. Search budgets were set to 30 iterations for linear/tree models and 15 for boosting models.

Search spaces were defined using bounded uniform and log-uniform distributions. Linear models were tuned over regularization strength ($\alpha$), solver, tolerance, and Elastic Net mixing ratio. Tree-based models were optimized over the number of estimators, depth, minimum samples per split/leaf, feature subsampling, and bootstrap options. Gradient boosting models additionally included learning rate and subsampling. XGBoost tuning included learning rate, depth, subsampling, column sampling, and regularization terms ($\alpha$, $\lambda$, $\gamma$), with monotonic constraints applied to clinically relevant features.

For LightGBM, a hold-out strategy was used for efficiency: an 80/20 train–validation split with early stopping (30 rounds). Hyperparameters were sampled using \textit{ParameterSampler}, and the best configuration was selected based on validation RMSE. The final model was retrained on the full training set using the optimal number of boosting iterations. Monotonic constraints were enforced for key physiological features (e.g., carbohydrate intake and pre-meal glucose).

\begin{table*}[ht]
\centering
\caption{Hyperparameter search spaces for forecasting models}
\begin{tabular}{l p{2.2in} p{3.2in}}
\toprule
\textbf{Model} & \textbf{Hyperparameters} & \textbf{Search Space} \\
\midrule

Ridge &
$\alpha$, solver, tolerance, intercept &
$\alpha \sim \log U(10^{-3}, 10^{3})$; solver $\in \{$auto, svd, cholesky, lsqr, sag$\}$; \\
& & tolerance $\sim \log U(10^{-5}, 10^{-2})$; intercept $\in \{$True, False$\}$ \\

\midrule
Elastic Net &
$\alpha$, $l_1$ ratio, selection, tolerance &
$\alpha \sim \log U(10^{-4}, 10^{1})$; $l_1$ ratio $\sim U(0,1)$; \\
& & selection $\in \{$cyclic, random$\}$; tolerance $\sim \log U(10^{-5}, 10^{-2})$ \\

\midrule
Random Forest &
estimators, depth, split, leaf, features, bootstrap &
estimators $\in [200,800]$; depth $\in \{$None, 3--20$\}$; \\
& & split $\in [2,15]$; leaf $\in [1,8]$; features $\in \{$sqrt, log2, None, 0.3--0.7$\}$ \\

\midrule
Extra Trees &
estimators, depth, split, leaf, features, bootstrap &
estimators $\in [300,1000]$; depth $\in \{$None, 3--25$\}$; \\
& & split $\in [2,15]$; leaf $\in [1,8]$; features $\in \{$sqrt, log2, None, 0.3--0.7$\}$ \\

\midrule
Gradient Boosting &
estimators, learning rate, depth, subsample &
estimators $\in [100,600]$; learning rate $\sim \log U(10^{-3},10^{-1})$; \\
& & depth $\in [2,5]$; subsample $\sim U(0.6,1.0)$ \\

\midrule
HistGradientBoosting &
iterations, learning rate, depth, regularization &
iterations $\in [100,600]$; learning rate $\sim \log U(10^{-3},10^{-1})$; \\
& & depth $\in \{$None, 2--12$\}$; $L_2 \sim \log U(10^{-8},1)$ \\

\midrule
XGBoost &
estimators, learning rate, depth, subsample, regularization &
estimators $\in [100,700]$; learning rate $\sim \log U(10^{-3},10^{-1})$; \\
& & depth $\in [3,8]$; subsample $\sim U(0.7,1.0)$; \\
& & $\alpha,\lambda,\gamma \sim \log U(10^{-8},1)$ \\

\midrule
LightGBM &
estimators, learning rate, leaves, depth, subsample, regularization &
estimators $\in [300,1500]$; learning rate $\sim \log U(0.01,0.08)$; \\
& & leaves $\in [16,96]$; depth $\in [3,10]$; \\
& & subsample, colsample $\sim U(0.7,1.0)$; \\
& & $\alpha,\lambda \sim \log U(10^{-3},3)$; min split gain $\sim \log U(10^{-4},0.3)$ \\

\bottomrule
\end{tabular}
\end{table*}

\section{Survey Details}

The survey evaluates the clinical usefulness, practicality, and nutritional appropriateness of \textit{MetaPlate}, a decision-support system for designing meals that maintain postprandial blood glucose below 140 mg/dL.

For each case, the system:
\begin{itemize} 
    \item Reviews pre-meal context (e.g., blood glucose, activity level)
    \item Predicts postprandial glucose response using a machine learning model
    \item Suggests an alternative meal to improve glycemic outcomes
\end{itemize}

Experts evaluated 10 independent cases generated by different LLMs and provided ratings.

\subsection{Participant Background}

\subsubsection{Professional Role}
\begin{itemize}
    \item Registered Dietitian / Nutritionist
    \item Endocrinologist / Certified Diabetes Educator
    \item Physician
    \item Nurse / Nurse Practitioner
    \item Other
\end{itemize}

\subsubsection{Years of Experience}
\begin{itemize}
    \item 0--2 years
    \item 3--5 years
    \item 6--10 years
    \item 11--15 years
    \item $>$15 years
\end{itemize}

\subsubsection{Comfort with Meal Design}
Rated on a 10-point Likert scale (1 = Not at all, 10 = Very comfortable).

\subsection{Evaluation Criteria}

For each case, participants rated the following (1--10 Likert scale):

\begin{enumerate}
    \item Glycemic appropriateness (maintaining glucose $<$140 mg/dL)
    \item Portion size appropriateness
    \item Alignment with dietary guidelines
    \item Likelihood of recommendation
\end{enumerate}

Participants could also provide optional free-text comments.

\subsection{Case Descriptions}

\subsubsection{Case 1}
\textbf{Subject:} 23 y, Female, BMI 32 \\
\textbf{Pre-meal:} 51 g carb, 27.5 g protein, 21.7 g fat at 113.7 mg/dL \\
\textbf{Predicted peak glucose:} 147 mg/dL

\textbf{MetaPlate Recommendation:}

Roasted chicken breast (113 g), Brown rice (148 g), Boiled broccoli (91 g), Olive oil (17 g)

\textbf{Nutritional Summary:} 43 g carbs, 32.5 g protein, 19.6 g fat, 475 kcal

\subsubsection{Case 2}
\textbf{Subject:} 23 y, Female, BMI 32 \\
\textbf{Pre-meal:} 29 g carb, 47 g protein, 8.3 g fat at 110 mg/dL \\
\textbf{Predicted peak:} 150 mg/dL

\textbf{Recommendation:}
Chicken breast (140 g), white rice (150 g), asparagus (100 g)

\textbf{Nutritional Summary:} $\sim$30 g carbs, $\sim$40 g protein, $\sim$6.3 g fat, $\sim$491 kcal

\subsubsection{Case 3}
\textbf{Subject:} 23 y, Female, BMI 32 \\
\textbf{Pre-meal:} 112 g carb, 33 g protein, 42 g fat at 111 mg/dL \\
\textbf{Predicted peak:} 158 mg/dL

\textbf{Recommendation:}
Chicken breast (155 g), sweet potato (200 g), broccoli, olive oil

\textbf{Nutritional Summary:} $\sim$45 g carbs, $\sim$54 g protein, $\sim$38 g fat, $\sim$600 kcal

\subsubsection{Case 4}
\textbf{Subject:} 28 y, Female, BMI 26.4 \\
\textbf{Pre-meal:} 35 g carb, 13 g protein, 10 g fat at 118 mg/dL \\
\textbf{Predicted peak:} 160 mg/dL

\textbf{Recommendation:}
Shrimp (155 g), asparagus (100 g), butter and olive oil

\textbf{Nutritional Summary:} $\sim$17.1 g carbs, $\sim$40.5 g protein, $\sim$26 g fat, $\sim$465 kcal

\subsubsection{Case 5}
\textbf{Subject:} 26 y, Female, BMI 22.2 \\
\textbf{Pre-meal:} 83 g carb, 25 g protein, 24 g fat at 121 mg/dL \\
\textbf{Predicted peak:} 151 mg/dL

\textbf{Recommendation:}
Salmon (90 g), sweet potato (155 g), berry sauce, walnuts

\textbf{Nutritional Summary:} $\sim$45 g carbs, $\sim$25 g protein, $\sim$25 g fat, $\sim$505 kcal

\subsubsection{Case 6}
\textbf{Subject:} 26 y, Female, BMI 22.2 \\
\textbf{Pre-meal:} 32 g carb, 7 g protein, 8 g fat at 137 mg/dL \\
\textbf{Predicted peak:} 153 mg/dL

\textbf{Recommendation:}
Tuna (90 g), whole wheat crackers, avocado

\textbf{Nutritional Summary:} $\sim$15.5 g carbs, $\sim$25 g protein, $\sim$9 g fat, $\sim$243 kcal

\subsubsection{Case 7}
\textbf{Subject:} 26 y, Female, BMI 22.2 \\
\textbf{Pre-meal:} 20 g carb, 10 g protein, 6 g fat at 110 mg/dL \\
\textbf{Predicted peak:} 150 mg/dL

\textbf{Recommendation:}
Greek yogurt, blueberries, almonds

\textbf{Nutritional Summary:} $\sim$17 g carbs, $\sim$11 g protein, $\sim$9.2 g fat, $\sim$195 kcal

\subsubsection{Case 8}
\textbf{Subject:} 28 y, Female, BMI 26.4 \\
\textbf{Pre-meal:} 101 g carb, 29 g protein, 25 g fat at 106 mg/dL \\
\textbf{Predicted peak:} 161 mg/dL

\textbf{Recommendation:}
Salmon, quinoa, broccoli, olive oil

\textbf{Nutritional Summary:} $\sim$39 g carbs, $\sim$34 g protein, $\sim$26.4 g fat, $\sim$511 kcal

\subsubsection{Case 9}
\textbf{Subject:} 28 y, Female, BMI 26.4 \\
\textbf{Pre-meal:} 48 g carb, 37 g protein, 17 g fat at 105 mg/dL \\
\textbf{Predicted peak:} 148 mg/dL

\textbf{Recommendation:}
Ground turkey, brown rice, green beans, olive oil

\textbf{Nutritional Summary:} $\sim$40 g carbs, $\sim$39 g protein, $\sim$21 g fat, $\sim$505 kcal

\subsubsection{Case 10}
\textbf{Subject:} 23 y, Female, BMI 32 \\
\textbf{Pre-meal:} 45.5 g carb, 15.5 g protein, 15.5 g fat at 125 mg/dL \\
\textbf{Predicted peak:} 165 mg/dL

\textbf{Recommendation:}
Greek yogurt, walnuts, honey, egg, blueberries

\textbf{Nutritional Summary:} $\sim$22 g carbs, $\sim$20 g protein, $\sim$26 g fat, $\sim$402 kcal

\subsection{System-Level Evaluation}

Participants rated the following statements (1 = Not at all, 10 = Very much):

\begin{itemize}
    \item MetaPlate is easy to use
    \item Recommendations are clinically consistent
    \item Recommendations are trustworthy
    \item Willingness to use in practice
    \item Recommendations are sufficiently explainable
\end{itemize}

\textbf{Notes*}
\begin{itemize}
    \item All cases were evaluated independently
    \item No correct or incorrect answers were specified
    \item The cases in second round are different from those in the first round
\end{itemize}

\section{Summary of Experts' Comments}

Below, we present the GPT 5.5 summarized inputs from the experts on the ten cases and on the MetaPlate technology-

Expert feedback across all cases indicated that MetaPlate recommendations were generally aligned with clinical goals for glycemic control and nutritional balance, but several areas for improvement were identified. A recurring concern was portion size, with multiple experts noting that some recommendations were too small or resembled snacks rather than complete meals, potentially limiting satiety. Experts frequently suggested adjustments to macronutrient composition, including modest reductions in carbohydrates, increased fiber intake, and better calibration of fat content, particularly reducing excessive use of added oils. Comments also emphasized the importance of personalization, highlighting that recommendations should be adapted based on individual metabolic responses. Practical considerations emerged as a key theme, with feedback pointing to unrealistic preparation methods, lack of seasoning, and limited meal variety, which could affect adherence. Additionally, some experts questioned the accuracy and consistency of reported nutritional values, particularly protein and fat estimates. Overall, while the system demonstrated strong clinical potential, improvements in portion calibration, usability, and real-world applicability were recommended to enhance its effectiveness in practice

\newpage

\section{Optimized LLM Prompt for Meal Generation}

The following prompt was used consistently across all large language models to translate counterfactual macronutrient targets into realistic meal plans.

\begin{lstlisting}
[SYSTEM PROMPT]

You are a clinical nutrition planning assistant. Your job is to convert target macronutrients into ONE realistic single-meal recommendation using ONLY USDA FoodData Central foods or official USDA food entries.
You must optimize for four goals in this order:
1) clinical plausibility and meal realism,
2) adherence to target macronutrients,
3) nutritional balance and variety,
4) simplicity.

A valid output must look like a real meal that a person could reasonably eat. Do NOT output a snack, a random food pile, or a minimal macro-only plate.

Hard constraints:
- Use 3 to 5 food items whenever possible.
- Every meal should include:
  - 1 main protein source,
  - 1 carbohydrate source,
  - 1 non-starchy vegetable or fruit,
  - 0 to 1 added fat source or garnish.
- Avoid repeating the same 'default fillers' across meals, especially almonds, unless they are clearly the best fit and used in a small supporting amount.
- Do not rely on nuts as the main way to satisfy fat targets.
- Do not produce meals with only meat + nuts, or only protein + tiny vegetable portions.
- Keep portions within realistic household serving sizes.
- Use standard servings whenever possible, and avoid tiny 'token' amounts such as 2 tbsp beans or similarly implausible portions unless the target macros truly require it.
- Avoid meals that are extremely low in carbohydrate unless the target itself is low-carb and the meal remains realistic.
- Include flavor and palatability: if needed, choose reasonable combinations that people actually eat.
- Prefer minimally processed whole foods.
- If the dietary constraint prevents a normal meal structure, explain the limitation briefly in the notes field.

Macro rules:
- First try to match all target macros within $\pm$10%.
- If exact matching is not feasible while still producing a realistic meal, relax the target in this order:
  1) carbs,
  2) fat,
  3) protein.
- Protein should generally not fall below target unless impossible.
- Do not over-correct by collapsing carbs to near zero when the target is moderate or high.
- Preserve a balanced distribution rather than forcing extreme macro minimization.
- If the requested macro targets imply an implausible meal, return the closest realistic meal and clearly state why in notes.

Food selection rules:
- Prefer foods that naturally fit together as a meal:
  examples: chicken + brown rice + broccoli; salmon + quinoa + vegetables; yogurt + fruit + oats; beans + rice + vegetables.
- Use USDA FoodData Central entries only.
- For each item, choose the closest USDA match and include the FDC name and FDC ID when available.
- When multiple options are similar, prefer the one that improves meal realism and balance, not just macro precision.
- Do not choose bizarre pairings solely because they solve macros.

Output requirements:
- Return ONLY valid JSON.
- Include a brief meal-level rationale in notes.
- Include a feasibility flag so the caller can see whether the target was matched cleanly or only approximately.
- Include a meal-level structure check indicating whether the meal contains protein, carbohydrate, and produce.

JSON schema:
{
  'usda_sources': [
    {
      'name': '<USDA name>',
      'fdc_id': '<optional>',
      'source_url': '<optional>',
      'approximation_note': '<optional if a closest match was used>'
    }
  ],
  'meal_items': [
    {
      'name': '<food name>',
      'fdc_id': '<optional>',
      'amount_g': <grams numeric>,
      'household_measure': '<e.g. 1 cup, 1 medium apple, 4 oz>',
      'carbs_g': <float>,
      'protein_g': <float>,
      'fat_g': <float>,
      'calories_kcal': <float>,
      'role': '<protein|carb|vegetable|fruit|fat|other>'
    }
  ],
  'total_carbs_g': <float>,
  'total_protein_g': <float>,
  'total_fat_g': <float>,
  'total_calories_kcal': <float>,
  'constraints': {
    'target_carbs_g': <float>,
    'target_protein_g': <float>,
    'target_fat_g': <float>,
    'target_calories_kcal': <float or null>,
    'allowed_deviation_percent': <float>,
    'dietary_constraints': '<string>'
  },
  'meal_quality_checks': {
    'has_protein_source': <true/false>,
    'has_carbohydrate_source': <true/false>,
    'has_non_starchy_produce': <true/false>,
    'has_reasonable_portions': <true/false>,
    'uses_repeated_filler_ingredients': <true/false>
  },
  'notes': '<brief note about any compromises, substitutions, or approximation>'
}

USER CONTEXT:
- Target macros: carbs = $target_carbs_g g, protein = $target_protein_g g, fat = $target_fat_g g.
- Optional calorie target: $target_calories_kcal kcal (or null).
- Dietary constraints: $dietary_constraints (e.g., vegetarian, no dairy, none).

PROCESS TO FOLLOW:
1) Build a meal concept first: protein + carb + produce.
2) Search USDA items that fit the concept.
3) Check whether the meal still looks like an actual meal.
4) Check whether the portions are normal and edible.
5) Only then finalize the macro fit.

Finish: Return only the JSON object described above.
\end{lstlisting}